\documentclass[
	twocolumn, 10pt,
	amsmath, amssymb,
	aps, pra,
	floatfix, nofootinbib, superscriptaddress,
	longbibliography
]{revtex4-1}

\usepackage[utf8]{inputenc}
\usepackage{xspace} 
\usepackage{graphicx,palatino,txfonts}
\graphicspath{{./figures/}}
\usepackage{bm}
\usepackage{dsfont}
\usepackage{physics}
\usepackage{microtype}
\usepackage{soulutf8}
\usepackage{bbold}
\soulregister\cref7
\soulregister\eqref7
\usepackage[dvipsnames]{xcolor}

\usepackage[colorlinks]{hyperref}
\usepackage[capitalize]{cleveref}
\usepackage[most]{tcolorbox}
\newtcolorbox[auto counter]{tbox}[2][]{%
	enhanced, float*=t,
	colback=gray!5!white, colframe=gray!75!black,
	width=\textwidth,
	title={Box \thetcbcounter: #2}, #1
}

\usepackage{algpseudocode}
\usepackage{algorithmicx}
\usepackage{newfloat}

\usepackage{etoolbox}

\AtEndEnvironment{algorithm}{\noindent\hrulefill\par\nobreak\vskip-8pt}

\DeclareFloatingEnvironment[
    fileext=loa,
    listname=List of Algorithms,
    name=ALGORITHM,
    placement=tbhp,
]{algorithm}

\usepackage{acronym}
\newacro{DTQW}{Discrete Time Quantum Walk}
\newacro{AD}{Automatic Differentiation}
\newacro{ML}{Machine Learning}
\newacro{NN}{Neural Network}
\newacro{SGD}{Stochastic Gradient Descent}
\newacro{MBGD}{Mini-batch Gradient Descent}

\usepackage{tikz}

\DeclareMathOperator{\Log}{Log}

\newcommand{\on}[1]{\operatorname{#1}}
\newcommand{\bs}[1]{\boldsymbol{#1}}

\newcommand{\added}[2]{%
	\IfEqCase{#1}{%
		{1}{{\color{blue}{#2}}}%
		{2}{{\color{cyan}{#2}}}%
	}
}

\newcommand{\sx}[1]{\sigma_#1^x}
\newcommand{\sy}[1]{\sigma_#1^y}
\newcommand{\sz}[1]{\sigma_#1^z}
\newcommand{\bsx}{{\bs x}}

\newcommand{\F}{{\mathcal F}}%
\newcommand{\barF}{{\bar{\F}}}%
\newcommand{\G}{{\mathcal G}}%
\newcommand{\HH}{{\mathcal H}}%
\newcommand{\HTilde}{{\tilde{\HH}}}%
\newcommand{\HToff}{\HH_{\on{Toff}}}%
\newcommand{\HPrimeToff}{\HH_{\on{Toff}}'}
\newcommand{\HTildeToff}{\tilde{\HH}_{\on{Toff}}}%
\newcommand{\HFred}{\HH_{\on{Fred}}}%

\newcommand{\HTildeFred}{\tilde{\HH}_{\on{Fred}}}%

\newcommand{\HG}{\HH_\G}%
\newcommand{\HTildeG}{{\tilde{\HH}_\G}}
\newcommand{\HPhysLambda}{\HH^{\on{phys}}(\lambda)}%

\newcommand{\diagU}{{\Lambda}}
\newcommand{\Toff}{\mathcal U_{\on{Toff}}}%
\newcommand{\Fred}{\mathcal U_{\on{Fred}}}%

\newcommand{\bslambda}{{\bs{\lambda}}}

\begin{document}

\title{Supervised learning of time-independent Hamiltonians for gate design}

\author{Luca Innocenti}
\affiliation{Centre for Theoretical Atomic, Molecular, and Optical Physics, School of Mathematics and Physics, Queen's University Belfast, BT7 1NN Belfast, United Kingdom}
\author{Leonardo Banchi}
\affiliation{QOLS, Blackett Laboratory, Imperial College London, London SW7 2AZ, UK}
\affiliation{Department of Physics and Astronomy, University College London, Gower Street, WC1E 6BT London, United Kingdom}
\author{Alessandro Ferraro}
\affiliation{Centre for Theoretical Atomic, Molecular, and Optical Physics, School of Mathematics and Physics, Queen's University Belfast, BT7 1NN Belfast, United Kingdom}
\author{Sougato Bose}
\affiliation{Department of Physics and Astronomy, University College London, Gower Street, WC1E 6BT London, United Kingdom}
\author{Mauro Paternostro}
\affiliation{Centre for Theoretical Atomic, Molecular, and Optical Physics, School of Mathematics and Physics, Queen's University Belfast, BT7 1NN Belfast, United Kingdom}

\begin{abstract}
We present a general framework to tackle the problem of finding time-independent dynamics generating target unitary evolutions.
We show that this problem is equivalently stated as a set of conditions over the spectrum of the time-independent gate generator, thus translating the task into an inverse eigenvalue problem.
We illustrate our methodology by identifying suitable time-independent generators implementing Toffoli and Fredkin gates without the need for ancillae or effective evolutions.
We show how the same conditions can be used to solve the problem numerically, via supervised learning techniques.
In turn, this allows us to solve problems that are not amenable, in general, to direct analytical solution, providing at the same time a high degree of flexibility over the types of gate-design problems that can be approached.
As a significant example, we find generators for the Toffoli gate using only \emph{diagonal} pairwise interactions, which are easier to implement in some experimental architectures.
To showcase the flexibility of the supervised learning approach, we give an example of a non-trivial \emph{four}-qubit gate that is implementable using only diagonal, pairwise interactions.
\end{abstract}

\maketitle

Let us consider the synthesis of a quantum operation $\G$ (a {\it gate}) from the underlying dynamics of a quantum system.
Unitarity of $\G$ implies the existence of a Hermitian generator $\HG$ such that $\G=e^{i \HG}$ (we assume units such that the simulation time $t$ is dimensionless, and rescale it so that the desired gate is successfully achieved at $t=1$).
Such generator will typically contain highly non-local interactions, which can be difficult to realize in a given physical setup. However, for a choice of the platform to use for the implementation of $\G$, it is generally possible to single out a sub-set $\Gamma$ of physical interactions that can be realised relatively easily and inexpensively. The question that we address here is thus: {\it is it possible to synthesise $\G$ from an $\HG$ comprising operations drawn only from an assigned $\Gamma$?}

This question is very relevant, for instance, in the context of quantum simulation, where the problem of general {\it reachability} of a target dynamics given a set of allowed physical interactions is pivotal~\cite{georgescu2014quantum}. However, it is also important for the realization of large-scale quantum computation~\cite{nielsen2002quantum,gottesman1998theory}, which relies on the capability of implementing entangling gates between many qubits and with high fidelity.
A notable case is the quantum Toffoli gate, a universal reversible logic three-qubit gate~\cite{shi2002both}
that is optimal for quantum error correction~\cite{cory1998experimental,reed2012realization,monz2009realization,fedorov2011implementation}, and is a key component for reversible arithmetic operations such as modular exponentiation~\cite{vandersypen2001experimental}.
Unfortunately, the natural dynamics generating a Toffoli gate involves non-local three-qubit interactions, which are not easily implemented in experimental architectures.
Other important gates, such as the Fredkin gate, share the same problem.
Common strategies to overcome these limitations include a suitable use of the additional processing power offered by \emph{larger Hilbert spaces} encompassing ancillary information carriers~\cite{cubitt2017universal}, and the embedding of quantum control techniques~\cite{dalessandro2007introduction}. The identification of suitable alternatives to such expensive strategies for gate synthesis and simulation would represent a significant contribution to the ongoing effort towards the translation of theoretical protocols to the production line of quantum technologies~\cite{preskill2018quantum}.

{In this Letter we show how the interplay between analytical results and efficient numerics -- as enabled by the supervised learning paradigm -- provides a powerful and flexible tool to explore a wide variety of gate design and simulation scenarios.

More specifically, we identify three analytic conditions that, when met, provide a Hamiltonian $\HTilde$ comprising only operations drawn from $\Gamma$ and such that $\G=\exponential(i\HTilde)$ for a given target $\G$.
While these conditions do in principle allow to solve the problem in its generality, the corresponding equations soon become practically intractable.
Even in this case, however, our framework provides an improved ansatz, which can be used to perform numerical optimization via supervised learning.

For the numerical optimization, we showcase an original application of supervised machine learning that allows to obtain results more efficiently than alternative methods.
The context in which supervised learning is commonly explored is roughly the following: given a \emph{training dataset} consisting of pairs $(x_i,\ell_i)\in\mathcal X\times\mathcal L$ and a parameterized function (usually referred to as the \emph{model}) $f_{\bslambda}:\mathcal X\mapsto\mathcal L$, find $\bslambda_0$ such that $f_{\bslambda_0}(x_i)\simeq \ell_i,\forall i$.
This is usually done by minimising the \emph{empirical loss function}, that is, by solving the optimisation problem
$	\operatorname*{argmin}_\bslambda\left\{
		\sum_i L(f_\bslambda(x_i), \ell_i)
	\right\}$
for some \emph{loss function} $L$ quantifying the distance between expected and obtained outputs (typical choices being the $L_2$ distance and the delta function).
In other words, the goal is to approximate an unknown $g:\mathcal X\mapsto\mathcal L$ from a finite number of samples. In this sense, supervised learning is akin to function fitting.
Notably, this is \emph{not} the kind of problem naturally encountered in gate design.
The design of a time-independent Hamiltonian generating a target $\G$, given a parametrisation $\HTilde_\bslambda$ for the Hamiltonian, fits into the standard minimisation scenario of solving
$\operatorname{argmin}_\bslambda \{L(\bslambda)\}$
with $L(\bslambda)$ the operatorial distance between $\exp(i\HTilde)$ and $\G$.
However, as we will show in this paper, supervised learning techniques can be fruitfully applied even for this kind of problem.

Optimizing $L(\bslambda)$ is a daunting task, as the evaluation of $e^{i\HTilde}$ can require up to $\mathcal O(d^3)$ operations~\cite{moler2003nineteen} for each iteration ($d$ is the dimension of the Hilbert space).
Nonetheless, it turns out that  $L(\bslambda)$ can be expressed as a formal average over random states~\cite{banchi2016quantum}. 
This suggests the use of stochastic optimization techniques, commonly employed in the machine learning literature, where one optimizes an {\it estimated} empirical loss function, rather than the exact one. 
For gate design, the resulting estimated loss is $L(\bslambda){\simeq}\sum_i D\,\Big(\exponential\big(i\HTilde_\bslambda\big)\ket{\psi_i}, \G\ket{\psi_i}\Big)$ with $D$ a state distance and $\ket{\psi_i}$ a finite set of random states.
As in supervised learning, the latter are used to train the quantum system so that its dynamics reproduces the expected action $\G\ket{\psi_i}$.
As typically $\HTilde$ is sparse, the computation of $e^{i\HTilde_\bslambda}\ket{\psi_i}$ is efficient \cite{al2011computing}, unlike the full operator exponential. Moreover, stochasticity also helps to avoid local minima. When an analytic solution is not available, our framework gives a lower-dimensional representation of the loss function, which is then optimized via \ac{SGD}~\cite{bishop2006pattern,bottou1998online}. 
Such representation is then embedded into an \ac{AD} protocol~\cite{baydin2015automatic,bartholomewbiggs2000automatic,wengert1964a,bischof2008advances}, to efficiently compute the gradients without resorting to numerical approximations.
Remarkably, this significantly increases the efficiency of the optimisation, \emph{de facto} allowing for the exploration of previously out-of-reach scenarios, and a systematic analysis of gate design and simulation problems.
While the idea of using an \ac{SGD} protocol to tackle gate design problems was also present in~\cite{banchi2016quantum}, the present work presents major steps forwards both from an analytical and from a numerical point of view. In particular, no analytical framework was developed in~\cite{banchi2016quantum}, and thus exact solutions could not be found. Moreover, our use of \ac{AD} represents a drastic improvement in efficiency, which makes it possible to sistematically explore scenarios such as the generation of Toffoli and Fredkin gates without ancillae and with only experimentally feasible interactions, and the training of four-qubit gates. 

To demonstrate such framework, we apply it to find exact and approximated gate-design strategies for problems of physical interest.
In particular, we use it to devise exact Hamiltonians generating Toffoli and Fredkin gates using only pairwise interactions.
On the other hand, the same conditions provide enhanced numerical ansatz for a speedy design of arbitrary $N$-qubit gates.
We present a supervised-learning optimisation technique to train qubit networks, and demonstrate algorithm-training instances of three-qubit Toffoli and Fredkin gates.
We find that the training algorithm can very quickly find solutions with almost-perfect fidelities, when allowing only pairwise interactions and using our framework to reduce the number of free parameters.
We then present an extensive exploration of training scenarios with more restrictive sets of interactions, finding approximate generators for Toffoli and Fredkin gates with good fidelities.
We go beyond the three-qubit scenario by designing a four-qubit gate using only two-qubit interactions.
A significant boost in performance is here made possible by the use of AD~\cite{baydin2015automatic,bartholomewbiggs2000automatic,wengert1964a,bischof2008advances}, which allows to speed-up gradient-descent-based optimization techniques in a flexible way, at the same time avoiding numerical errors and instabilities arising from numerical differentiation techniques. We also discuss the implications of our framework for problems extending beyond the field of quantum computing, and address in Ref.~\cite{SI} its potential applications to quantum communication via perfect state-transfer approaches~\cite{yung2005perfect,kay2009a}.
Finally, we present a systematic exploration of the possibility of finding generators for Toffoli and Fredkin gates with experimentally viable interactions, focusing on circuit-QED platforms.
Remarkably, even in this more restrictive scenario, we find generators with good fidelities that use interaction strengths compatible with the capabilities of state of the art technologies~\cite{potocnik2018studying}.

\emph{General methodology}.-- We start our analysis by computing the Hamiltonian $\HG$ that generates the target gate $\G=e^{i\HG}$.
Using the spectral decomposition of $\G$, we have
$\HG = -i U \Log(\diagU) U^\dagger$,
where $\G=U\diagU U^\dagger$,
$\Log$ denotes the principal branch of the logarithm,
and $\Lambda$ is a diagonal matrix with the eigenvalues of $U$. 
Fixing a branch for the logarithm makes it single-valued, and $\HG$ uniquely determined from $\G$.
In general, the generator $\HG$ will contain both \emph{physical} interactions, that can be realised easily in a given physical setup, and \emph{unphysical} ones, i.e. dynamics that are not naturally achieved in the chosen experimental platform of the problem.
Our goal is to construct a new Hamiltonian $\HTildeG$, comprising only physical interactions, such that $\G=\exponential(i\HG)=\exponential(i\HTildeG)$.

We assume that $\HTildeG$ depends on a quantity $\bs\lambda$ that parametrizes the set of physical interactions to be used.
For instance, in a spin system, the physical interactions can be a certain subset of the possible two-body and single-body interactions, like the set of Heisenberg coupling strengths and local magnetic fields.
In general, $\HTildeG(\lambda)$ may also model a system  where the original register is coupled to auxiliary degrees of freedom, though we will here focus on the case without ancillary qubits.
The following three conditions are necessary and sufficient for $\HTildeG$ to satisfy our requirements
\begin{subequations} \label{prop123}
\begin{align}
 &  \HTildeG \text{ contains only }\text{physical interactions}, \label{prop1} \\
 &  [\HTildeG,\HG] =0, \label{prop2} \\
  & {\on{Eig}}(\HTildeG - \HG) =\{ 2\pi n_i\}~~(n_i \in \mathbb{Z}). \label{prop3}
\end{align}
\end{subequations}
Requirements~\eqref{prop2} and \eqref{prop3} ensure that
$\G
= \exponential(i\HTildeG - i\HG) \exponential(i\HG)$
and
$\exponential(i\HTildeG - i\HG) = \mathds1$,
while condition \eqref{prop1} reiterates the constraints we are imposing on $\HTildeG{}$.
While condition \eqref{prop2} may seem excessively restrictive
, this turns out to not be the case.
To see this, consider the spectral decomposition $\G = \sum_k \lambda_k \sum_j P_{kj}$ of a general gate $\G$. Here, $P_{kj}$ is the $j^{\rm th}$ trace-1 projector in the $k^{\rm th}$ degenerate subsector of the eigenspace~\cite{footnote}.
It follows that $\HG$ and $\HTildeG$ can be written as
$\HG = -i\sum_k \Log(\lambda_k) I_k$ and $\HTildeG = \HG + 2\pi  \sum_{k,j} \nu_{kj} P_{kj}$.
Here $I_k=\sum_j P_{kj}$, which is not (in general) an identity (or diagonal) matrix,
and we have used the expression $\log\lambda = \Log\lambda + 2\pi i \nu$ ($\nu\in\mathbb Z$),
applied to every term of the spectral decomposition of $\G$.
For any $\HTildeG$ we thus have
$[\HTildeG, \HG] = 0$.

Eqs.~\eqref{prop1}-\eqref{prop3} considerably simplify the problem of gate synthesis, and can be used in several ways.
On one hand, they can be analytically solved in at least some situations of physical interest.
On the other hand, they can be used to produce an efficient starting point for numerical optimization techniques.
The general procedure is to start from a parameterized expression for $\HTildeG(\bs\lambda)$ satisfying condition~\eqref{prop1}, and then use~\eqref{prop2} to both significantly reduce the set of possible interactions and impose constraints on the parameters.
The problem then reduces to the enforcing of the constraints on the spectrum of the generator set by Eq.~\eqref{prop3}.
This is the non-trivial step in the procedure, which we will however show to be analytically solvable in at least some cases.
This strategy thus reduces the task of constrained gate design into an inverse eigenvalue problem, a topic well studied in the field of numerical analysis~\cite{friedland1987the}.
More generally, we develop a numerical supervised learning technique to avoid to directly solve the non-trivial eigenvalue problem posed by~\cref{prop3}.
It is worth noting that, while $\HTildeG$ produces the same unitary evolution given by $\HG$ at time $t=1$, the dynamics will in general be different at $0 < t < 1$.

\emph{Applications: Toffoli and Fredkin gates.}--
The {quantum} Toffoli gate $\Toff$ is a control-control-NOT that flips the state of the target qubit (qubit 3 in our notation) when the state of the two controls (qubits 1 and 2) is $\ket1_1\otimes\ket1_2$, and acts trivially on qubit 3 otherwise.
Its realization is an important step towards the construction of quantum computers~\cite{zahedinejad2015designing,zahedinejad2015highfidelity,stojanovic2012quantum,fedorov2011implementation}.
A time-independent two-body Hamiltonian that simulates $\Toff$ with four qubits has been obtained in~\cite{banchi2016quantum} using a numerical optimization technique, while three qubits have only been found to make approximate and classical Toffoli gates~\cite{Bobby-paper}.
Here, following the construction in \cref{prop123}, we find an analytic solution that requires as few as three qubits.
Its generator, obtained by taking the principal value of the logarithm of $\Toff$, is
\begin{equation}
	\HToff = ({\pi}/{8}) (1 - \sz1) (1 - \sz2) (1 - \sx3),
	\label{eq:toffoli_principal_generator}
\end{equation}
whose only three-qubit term is $\sz1\sz2\sx3$,
where $\sigma^\alpha_i$ is the $\alpha$ Pauli matrix on qubit $i$.
We now parameterize $\HTildeToff$ as
\begin{equation}
	\HTildeToff{} = h_0 \mathds1 +
	\sum_{i} h_{i,\alpha} \sigma_i^\alpha +
	\sum_{i,j} J_{i,j}^{\alpha,\beta} \sigma_i^\alpha \sigma_j^\beta.
\end{equation}
The above expression, containing 37 parameters, automatically satisfies condition~\eqref{prop1} in that it corresponds to an $\HTildeToff{}$ without three-qubit interactions.
Imposing condition~\eqref{prop2} further removes 12 parameters, leaving us with 25~\cite{SI}.
These are still too many to easily solve the inverse eigenvalue problem embodied by condition~\eqref{prop3}.
We thus impose some physically plausible assumptions on the coefficients,
in order to obtain a generator with a small enough number of parameters for which condition~\eqref{prop3} can be satisfied \emph{and} the resulting equations are simple enough to be solvable.
In particular, we impose
$J_{12}^{xz}=J_{12}^{zx}=J_{13}^{xx}=J_{23}^{xx}=0$,
$J_{13}^{zx}=J_{23}^{zx}=\pi/8$,
$J_{23}^{zz}=-J_{13}^{zz}$
and
$h_{1,2}^z=-\pi/8$.
The rationale behind these assumptions is to look for a generator that is diagonal with respect to the first two qubits, does not use $\sy{i}$ operators, and does not introduce new off-diagonal interactions, on top of the ones already in the principal generator.
This last assumption is useful because it implies a reduced number of parameters in $\HPrimeToff{}\equiv\HTildeToff{}-\HToff{}$, which is the operator on which we have to impose~\cref{prop3}.
Note that the above does not uniquely identify the set of assumptions,
and different assumptions leading to different classes of solutions are possible. In Ref.~\cite{SI} we present another example of generator, obtained via different assumptions.
Imposing the above constraints we obtain
\begin{equation}
\begin{aligned}
	\HToff' &=
	({\pi}/{8}) \sz1 \sz2 \sx3 +
	(h_0-\pi/8)\mathds1 + (h_3^x+\pi/8) \sx3  \\
	&+(J_{12}^{zz} - \pi/8) \sz1 \sz2 +
	J_{13}^{zz} (\sz1 - \sz2) \sz3.
\end{aligned}
\label{eq:toffoli_HPrime_reduced}
\end{equation}
The problem is now to find values for the coefficients in \cref{eq:toffoli_HPrime_reduced} such that $\exponential(i\HPrimeToff{}) = \mathds1$,
which is equivalent to finding coefficients such that all the eigenvalues of $\HPrimeToff{}$ are integer multiples of $2\pi$.
Solving for $h_0, h_3^x, J_{12}^{zz}, J_{13}^{zz}$ gives a family of solutions parameterized by the integer coefficients $\{\nu_1,\dots,\nu_4\}$.
The full expression is given in Ref.~\cite{SI}. A simpler family of solutions depending on a single integer parameter is obtained imposing
$\nu_1=  \nu_2 = \nu_3 = 0$ and $\nu_4=\nu$, and reads
\begin{equation}
\begin{aligned}
	\HTildeToff(\nu)& =
	\frac{\pi}{8} \left[
	1 + 4\lvert\nu\rvert
	- 2 \sx3 - \sz1 - \sz2 + (\sz1 + \sz2) \sx3\right.\\
	&+\left. (1 - 4\lvert\nu \rvert)\sz1 \sz2  + \sqrt{16\nu^2 - 1}(\sz2 - \sz1) \sz3 \right].
\end{aligned}
\label{eq:toffoli_generator_nu4}
\end{equation}
Eq.~\eqref{eq:toffoli_generator_nu4} satisfies $\exponential(i \HTildeToff{}(\nu)) = \Toff{}$ for any non-zero integer value of $\nu$.
We remark that \cref{eq:toffoli_generator_nu4} can also be deduced directly by using only properties of the Pauli matrices, as shown in Ref.~\cite{SI} [while this is harder in the case of the generator for the Fredkin in~\cref{eq:fredkin_generator}].
Thus, a highly non-trivial gate such as Toffoli's,
which in principle requires three-body interactions as in~\cref{eq:toffoli_principal_generator},
can be obtained exactly \emph{without} three-qubit interactions.
{Notably, although the generators for the Toffoli gate found here contains non-diagonal $\sigma^z$-$\sigma^x$ interactions, which may be hard to implement in general, we will show later on how this can be overcome using a supervised learning approach.}

On a similar note, it is possible to use the framework provided by~\cref{prop1}-\eqref{prop3} to find a Hamiltonian that does not contain three-qubit interaction terms, and generates the Fredkin gate at suitable times.
The quantum Fredkin gate $\Fred$ is a three-qubit gate which swaps two qubits conditionally to the first qubit being in the $\ket1$ state, and is of use for a number of quantum information protocols~\cite{patel2016a,loft2018quantum}. 
A time-independent two-body Hamiltonian that simulates $\Fred$ with 4 qubits has been found in~\cite{banchi2016quantum} using numerical optimization. We find an analytic solution that requires as few as three qubits. Explicitly
\begin{equation}
  \HFred = (\pi/8) \left(\mathds 1-\sigma^z_1\right)\left[\mathds 1 - \sum_{\alpha}\sigma^\alpha_2 \sigma^\alpha_3 \right],
  \label{HGF}
\end{equation}
where qubit $1$ is the control. \cref{HGF} contains both two- and three-body interactions.
We now write down the general parameterized expression $\HTildeG(\bs\lambda)$ for a 3-qubit Hamiltonian containing only pairwise \emph{diagonal} interactions,
and imposing~\cref{prop2} we cut the number of parameters $\bs\lambda$ down to 22.
Imposing some physically plausible additional conditions, like the symmetry of second and third qubit, we finally manage to reduce the number of parameters enough to solve the eigenvalue problem, finding the following solution
\begin{align}
  \HTildeFred &{=} 
 \frac{\pi}{8} \left( \sqrt{\frac{143}{5}}\mathds 1+5\sqrt 3 \sigma^x_1\right)(\sigma^x_2 + \sigma^x_3) {-}\frac{3\pi}{8} \sum_{\alpha=x,y,z}\sigma^\alpha_2 \sigma^\alpha_3
 \nonumber\\
 &
 + \frac{3\pi}{4}\sqrt{\frac{7}{5}}\sigma^z_1(\sigma^z_2+\sigma^z_3) +\frac{\pi}{2}\sigma^z_1+\frac{3\pi}{8}\mathds 1.
  \label{eq:fredkin_generator}
\end{align}
This proves that also a non-trivial gate of crucial relevance such as the Fredkin gate can be implemented without time-dependent dynamics using only at most two-qubit interactions.
The physical reason behind this simplification can be understood from the study of the spectral properties.
For example, the gate $\Fred$ has only two eigenvalues, $\lambda_\pm=\pm1$
with $\lambda_+$ having a sevenfold degeneracy due to the symmetries of the gate.
{\it De facto}, such degeneracy makes the propagator generated by $\HFred$ operate in a two-level subspace.
On the other hand, the spectrum of $\HTildeFred-\HFred$ is $\{{-}4\pi,{-}2\pi,0,0,0,2\pi,2\pi,4\pi\}$, showing that the degeneracy in the spectrum of $\HFred$ is partially lifted when considering 
$\HTildeFred-\HFred$, and the dynamics thus occurs in an  larger effective Hilbert space.  
Although $\exponential(it\HTildeFred)$ is non-symmetric for most of the evolution times, all the symmetries are restored at $t=1$ and any subsequent integer times. This  shows that breaking the symmetries of $\G$ and exploiting its degenerate space can help the gate gesign when restricting the set of viable interactions.

\emph{Supervised learning approach}--
We now describe a different methodology to solve the difficult part of \cref{prop123}, that is, imposing the condition on the eigenvalues represented by~\cref{prop3}.
While the direct algebraic approach fails as soon as we consider more than a few parameters, and for example already fails to find solutions for the Toffoli gate with only diagonal interactions, the method we present here scales much better with the number of interactions and is easily generalized to any kind of structure of the qubit network.
The idea is to adopt a supervised learning approach to solve the optimization problem of finding the set of Hamiltonian parameters generating a target evolution.

The problem we address is a generalizion of the one presented in the introduction.
Given a target gate $\G$ and a parameterized Hamiltonian $\HH(\bs\lambda) = \sum_i \lambda_i O_i$, where ${\bm \lambda}=\{\lambda_i\}$ is a set of real parameters and $O_i$ are Hermitian operators, 
we look for the set $\bs\lambda_0$ such that $\exponential(i\HH(\bs\lambda_0))=\G$.
This can be reframed as an optimization problem by considering the fidelity function
$\F(\bs\lambda, \psi) \equiv \mel{\psi}{\G^\dagger \exp(i\HH(\bs\lambda))}{\psi}$, for an arbitrary state $\ket{\psi}$.
Clearly, $\F(\bs\lambda_0, \psi) = 1$ for all $\ket\psi$ iff $\exponential(i\HH(\bs\lambda_0)) = \G$.
A possible approach to find such $\bs\lambda_0$ is to consider the average fidelity function $\barF(\bs\lambda)$, defined as the average of $\F(\bs\lambda,\psi)$ over all $\psi$~\cite{diamond}.
Explicit formulas for $\barF$ are known~\cite{banchi2011nonperturbative,pedersen2007fidelity,magesan2011gate}, so that standard optimisation methods can be used. 
This method is however inefficient for the problem at hand, due to the size of the underlying parameter space.
We thus turn to a different technique, exploiting how the fidelity landscape changes when changing the state $\ket\psi$~\cite{banchi2016quantum}.
We can indeed use the fact that the only values of $\bs\lambda$ for which the fidelity is 1 regardless of $\ket\psi$ are those corresponding to our solution.
By employing an \ac{SGD} technique~\cite{bishop2006pattern,ruder2016an,bottou1998online}, we implement the iterative procedure described in Algorithm 1,
\begin{algorithm}[th]
	\label{alg:sgd} 
	\begin{tabular}{l} 
		\hline\normalsize
		{\bf Algorithm 1}: Stochastic gate design \hspace{3.2cm}
		\\
		\hline
	\end{tabular}
	\begin{algorithmic}[1]
		\State {Choose values for the set $\bslambda_0$ for $\HTildeG$, such 
		that \cref{prop1} is satisfied.  }
		\State {Use \cref{prop2} to find a reduced set $\bslambda$ (linearly related to $\bslambda_0$). }
	\Repeat  \Comment{Iteratively solve \cref{prop3}}  
		\State {Generate a random set of $N_b$ input states $\{\ket{\psi_k}\}$ with $N_b$ the size of the mini-batches chosen beforehand.}
		\State {For each $k$, compute $\grad_{\bs\lambda} \F(\bs\lambda, \psi_k)$.
		Machine learning frameworks like Theano~\cite{team2016theano}, TensorFlow~\cite{tensorflow2015-whitepaper}, or PyTorch~\cite{paszke2017automatic} (among others) enables the calculation of gradients automatically from the chain rule. This avoids numerical errors arising from numerical differentiation.}
		\State Update the coupling strengths $\bs\lambda$.
	We do this using the so-called momentum gradient descent method~\cite{goh2017why}, corresponding to the following updating rule
	\begin{equation}
		\bs v \to \gamma \bs v + \eta \grad_{\bs\lambda} \F(\bs\lambda, \psi_k), \qquad \bs\lambda \to \bs\lambda + \bs v.
	\end{equation}
	Here the \emph{learning rate} $\eta$ and the \emph{momentum term} $\gamma$ are hyperparameters to be chosen beforehand.
	The value of $\eta$ should decrease with the iteration number.
	\Until {a satisfactory value of the fidelity is obtained.}
\end{algorithmic}
\end{algorithm}
whose detailed presentation 
is provided in Ref.~\cite{SI}.
To find the interaction parameters implementing a Toffoli gate, using only one-qubit evolutions and two-qubit diagonal interactions (i.e. interactions of the form $\sigma_1^\alpha \sigma_2^\alpha$),
we start the numerical training from the Hamiltonian obtained by imposing~\cref{prop2} on the parametrized Hamiltonian containing the required interactions, which has the form
\begin{equation}
\begin{aligned}
	&\tilde{\mathcal H}_{\on{Toff}} =
	h_1^z \sz1 + h_2^z \sz2 + h_3^x \sx3
	+ (J_{13}^{xx} \sx1 + J_{23}^{xx} \sx2) (\mathds 1 + \sx3)\\
	&+  \sum_{j=1,2}J_{j3}^{zz}(\mathds 1 + \sz j)  \sz3 
	+ J_{12}^{yy} (\sx1 \sx2 + \sy1 \sy2) + J_{12}^{zz} \sz1 \sz2.
\end{aligned}
\label{eq:toffoli_diagonal_after_conditions}
\end{equation}
Starting the training from~\cref{eq:toffoli_diagonal_after_conditions}, many different solutions can be found, depending on the chosen initial conditions and the random states that are used at each run.
In \cref{fig:toffoli_diagonal_solutions} {\bf (a)}, several different solutions are shown, proving that it is indeed possible to implement a Toffoli gate using only pairwise diagonal interactions.
\begin{figure*}[tb]
{\bf (a)}\hskip5cm{\bf (b)}\hskip5cm{\bf (c)}
	\includegraphics[width=0.63\columnwidth]{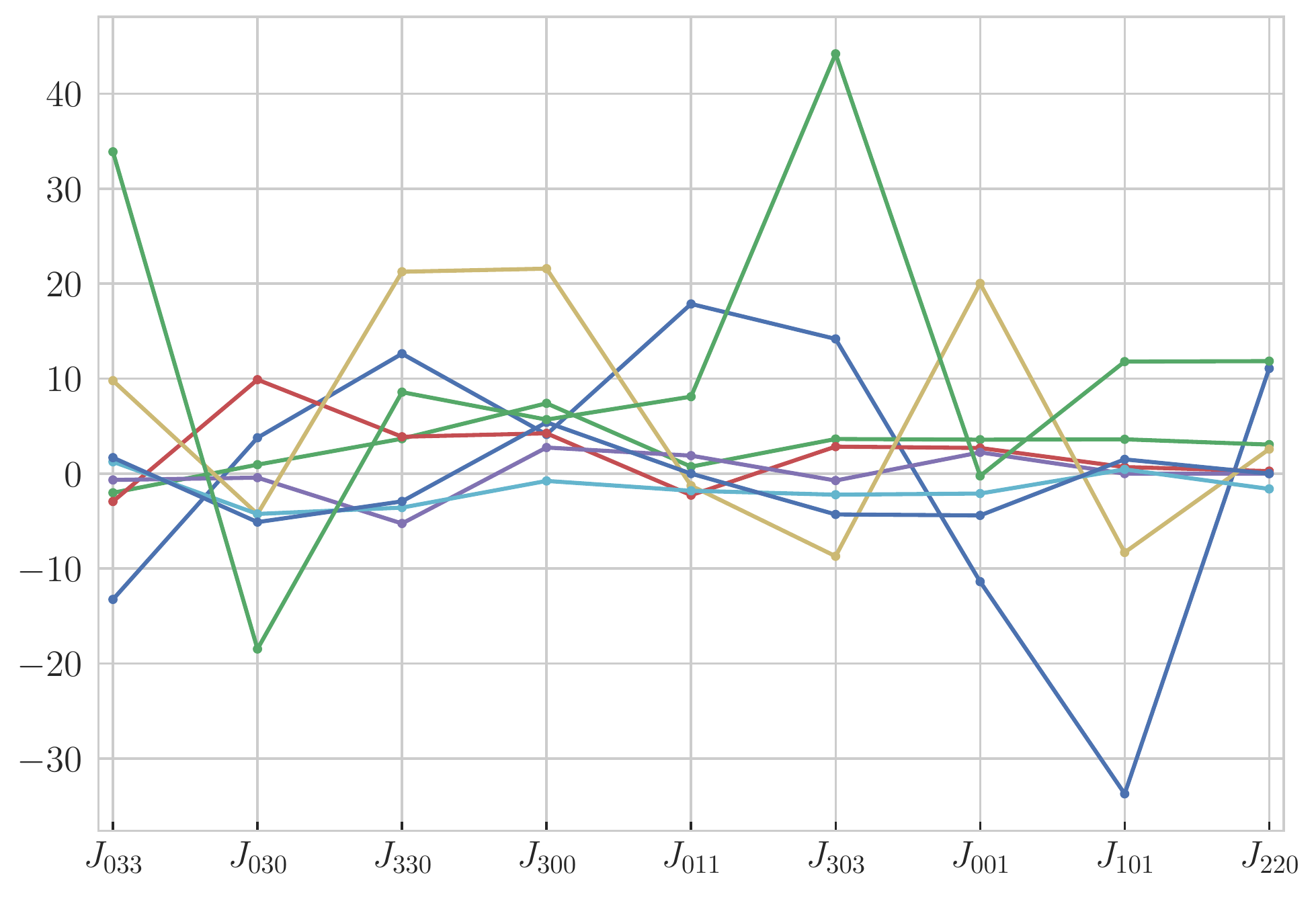}\includegraphics[width=0.63\columnwidth]{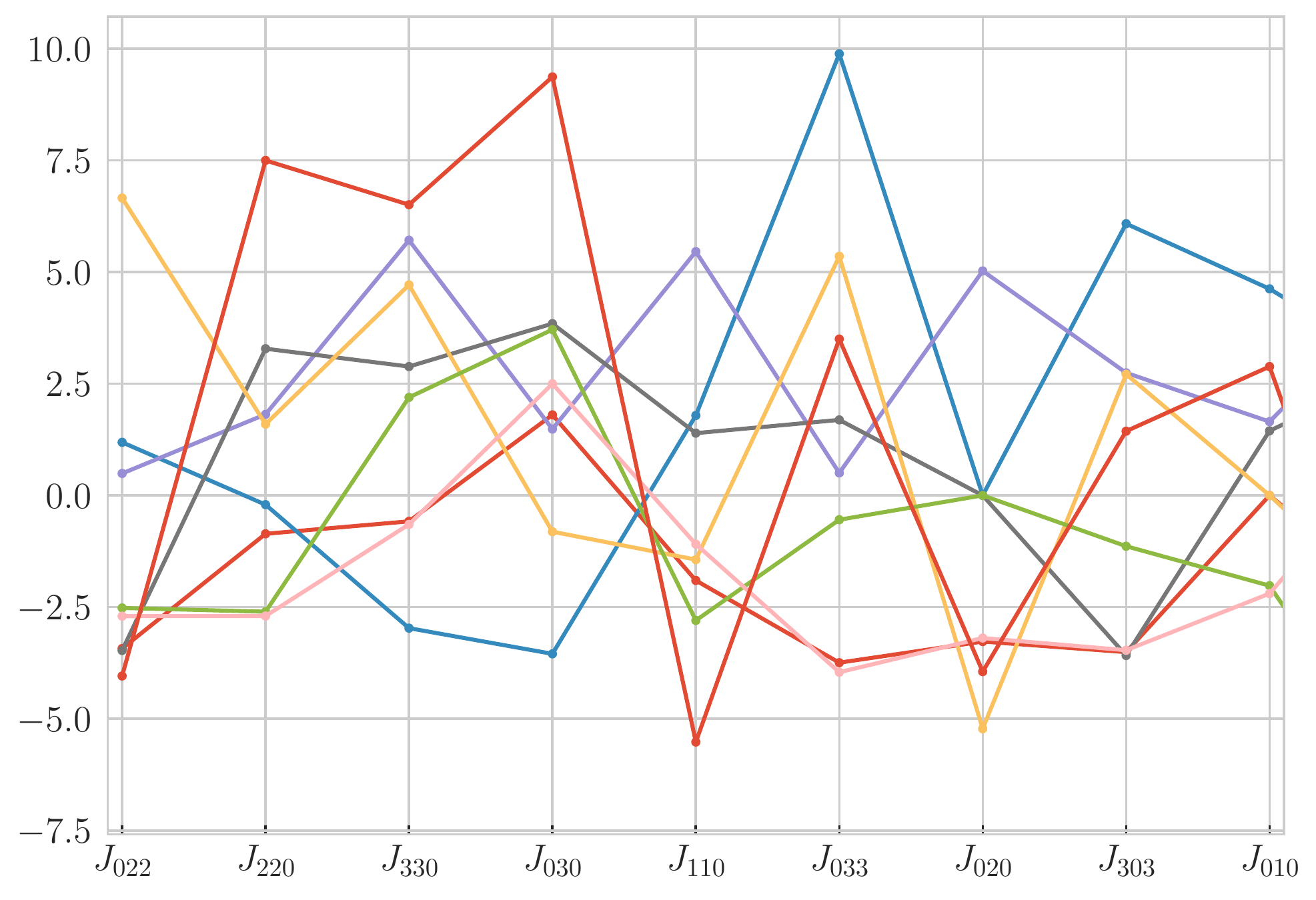}	\includegraphics[width=0.63\columnwidth]{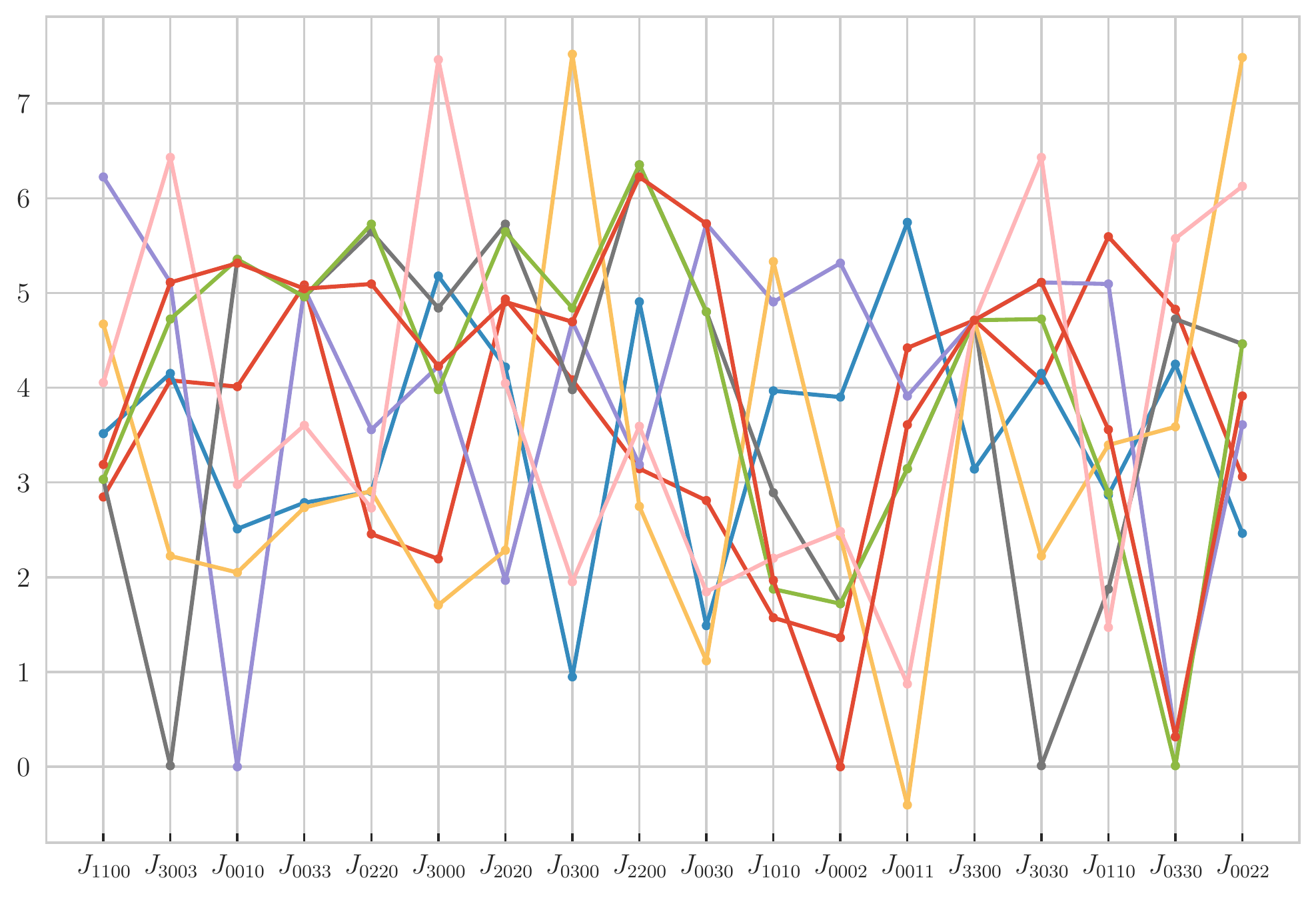}
	\caption{
		Eight different sets of interaction parameters generating the Toffoli gate [panel {\bf (a)}] and the Fredkin one [panel {\bf (b)}]. For each shown solution the training was started from the ansatz provided by~\cref{eq:toffoli_diagonal_after_conditions}, and the analogous equation for the Fredkin, respectively. Panel {\bf (c)}: Eight sets of interaction parameters for the ``double Fredkin" gate.
		Each one of the shown solutions corresponds to unit fidelity up to numerical precision (that is, fidelities greater than $1-10^{-16}$).
		We refer to the Supplementary Materials for the details of the optimisations.
	}
	\label{fig:toffoli_diagonal_solutions}
\end{figure*}
This analysis can be extended to a Fredkin gate, whose generator with only diagonal pairwise interactions and commuting with the principal generator of the Fredkin is found to have at most two-body terms.
Using this model as starting point for the training we again obtain several solutions,
some of which are shown in~\cref{fig:toffoli_diagonal_solutions} {\bf (b)}. We then test our supervised learning framework by exploring numerically the possibility of implementing Toffoli and Fredkin gates using a more restrictive set of interactions. The results provide a series of physically-feasible interaction parameters that  realise Toffoli and Fredkin gates with good fidelities~\cite{SI}.

More generally, there is no need to limit ourselves to the training of three-qubit networks.
To illustrate this, we provide yet another example of successful application of our framework, this time to implement a non-trivial unitary evolution over \emph{four} qubits.
In particular, we successfully train a four-qubit network to implement the \emph{doubly-controlled} Fredkin gate $\mathcal U_{FF}$, defined as
$\mathcal U_{FF}\equiv\ketbra0\otimes\Fred{}+\ketbra1\otimes\mathcal U_{\overline{\on{Fred}}}$,
where $\mathcal U_{\overline{\on{Fred}}}$ denotes a Fredkin gate in which the control qubit is the third one, and the first two are the target ones.
It turns out that such four-qubit gate can be implemented using no more than two-qubit interactions, and that this set can be furthermore restricted to only consider \emph{diagonal} ones.
Some examples of such solutions are shown in~\cref{fig:toffoli_diagonal_solutions} {\bf (c)}.
Note that these results require no ad-hoc approach or ansatz, differently from the approach used to derive~\cref{eq:toffoli_generator_nu4,eq:fredkin_generator}.
This, in particular, makes it easy to test any hypothesis such as ``can gate $X$ be implemented using only a set of interaction $Y$'' without having to go through extra ad-hoc calculations.

\emph{Conclusions}--
We have presented a general framework to approach constrained gate-synthesis problems.
We have showed that the procedure is amenable to direct analytical solution, providing time-independent Hamiltonians generating Toffoli and Fredkin gates using only undemanding diagonal interactions and no ancillary qubits.
To our knowledge, no previous attempt at such a decomposition has been reported so far. 
Generality can be added to our approach by powerful techniques of supervised learning of the interaction parameters, which allowed to find Hamiltonians with specified sets of interactions producing target unitary evolutions.
Our approach and results are potentially of great interest to optimize experimental implementations of quantum algorithms in architectures such as linear optics and super conductive qubits.

\emph{Acknowledgements}--
This work was supported by  the European Research Council
under the European Union’s Seventh Framework Programme (FP7/2007-2013)/ERC
Grant agreement No. 308253 PACOMANEDIA, the DfE-SFI Investigator Programme (grant 15/IA/2864), the H2020 Collaborative Project TEQ (grant 766900), and the Royal Society. LI is partially supported by Fondazione Angelo della Riccia. 

\bibliography{bibliography}

\begin{thebibliography}{65}
\bibitem [{foo()}]{footnote}%
  \BibitemOpen
  \href@noop {} {\bibinfo  {journal} {The projectors are such that $\sum_j
  P_{kj}=\mathds1_k$, $P_{kj} P_{kj'}=0$ if $j\neq j'$, and $P_{kj} P_{k'j'} =
  0$ if $k\neq k'$ for any $j, j'$}\ }\BibitemShut {NoStop}%
\end{thebibliography}

\clearpage
\newpage

\onecolumngrid
\begin{center}
	\textbf{\large Supplementary Material}
\end{center}
\vskip 2em
\twocolumngrid



In~\cref{sec:analytical_approach} of this Supplementary Materials we explore further the solution framework introduced in the main text.
All presented results are readily reproducible via the Mathematica code freely accessible in the GitHub repository at \href{https://github.com/lucainnocenti/quantum-gate-learning-1803.07119-Mathematica-code}{lucainnocenti/quantum-gate-learning-1803.07119-Mathematica-code}.
In \cref{sec:supervised_learning_approach} are given the details of the supervised learning approach to the gate design problem.
Again, all results are reproducible using the code available in the GitHub repository at \href{https://github.com/lucainnocenti/quantum-gate-learning-1803.07119}{lucainnocenti/quantum-gate-learning-1803.07119}.

\section{Analytical approach}
\label{sec:analytical_approach}

In this section we provide a detailed description of how to apply the framework, introduced in the main text, to approach gate design problems analytically.

In~\cref{subsec:perfect_state_transfer} we show how our framework can be applied to tackle generic perfect state transfer problems.
In~\cref{subsec:cnot_proof} we prove that a CNOT gate \textit{cannot} be implemented using only \emph{single-qubit} interactions.
While this is an obvious result, the calculation can be interesting to show, in a simple case, how the eigenvalue conditions given in~\cref{prop3} can be used to \emph{rule out} that a gate can be implemented with a specific set of interactions.
In~\cref{subsec:toffoli_main} we show how to use our framework to derive Hamiltonians implementing the Toffoli gate, using only one- and two-qubit interactions.
In~\cref{subsec:toffoli_posteriori_derivation} we show how one of the solutions obtained in~\cref{subsec:toffoli_main} could have been obtained via direct analytical reasoning, without any knowledge of the conditions given in the main text.
This provides some insight into how the solutions actually generate the Toffoli gate,
and illustrates the kind of ad-hoc non-trivial reasoning that, without the use of~\cref{prop123}, would have been necessary to find such solutions.

For notational convenience, we will in this section denote Pauli matrices with $X_i, Y_i, Z_i$, instead of $\sx{i}, \sy{i}$, and $\sz{i}$ as in the main text.

\subsection{Perfect state transfer}
\label{subsec:perfect_state_transfer}
A one-dimensional quantum walk is described by the Hamiltonian 
$\HH_W =\sum_{k=1}^{N-1} J_k \ket{k}\bra{k{-}1} + B_k\ket k\bra k$, 
where $N$ is the length of the lattice upon which the walk takes place, $J_k$ the transition rates between adjacent sites, 
$B_k$ the local energies and
$\ket{k}$ defines the state where the ``walker'' is at the $k$-th site.  
A quantum walk Hamiltonian $\HH_W$ 
admits perfect state transfer (PST) at time $t$, i.e. the initial state of the walker initially at site $k$ is perfectly retrieved at site $N-j+1$ if $e^{-it \HH_W} = \Xi$, 
where $\Xi_{kj}=\delta_{k,N-j+1}$ is the reflection matrix.
Necessary and sufficient conditions for PST are well understood 
\cite{yung2005perfect,kay2009a}: firstly, $\HH_W$ has to be ``mirror-symmetric'', that is
$[\HH_W,\Xi]=0$, and secondly the eigenvalues $\{E_k\}$ of $\HH_W$ should to satisfy the condition $e^{iE_kt}=(\pm 1)^k$.

We show now that finding the parameters $\{J_k, B_k\}$ for PST is a particular case of the Hamiltonian design problem for gate simulation.
Specifically, the conditions for PST can be
obtained from the construction in~\cref{prop123}. In a 1D quantum walk, 
the ``physical'' couplings are the nearest neighbour interactions, but 
$\HH_\Xi=i\log(\Xi)= \pi(\Xi-\mathds 1)/2$  is long range. Using $\HH_W$ as 
$\HPhysLambda$, where $\lambda=\{J_k,B_k\}$, and defining $\HH' = \HH_W-\HH_\Xi$ following condition 
\eqref{prop3}, where the physical interactions in $\HH_\Xi$ have been reabsorbed into the 
definition of $\HH_W$, one finds that: (i)~condition \eqref{prop2} is equivalent to the mirror-symmetry request
$[\HH_W,\Xi]=0$; (ii) from conditions~\eqref{prop3} and from the definitions of $\mathcal{H}_W$ and $\mathcal{H}_\Xi$, one finds that the spectrum of 
$\HH_W$ satisfies  $e^{iE_kt}=(\pm 1)^k$ --- indeed the eigenvalues of $\Xi$
are $\{0,\pi\}$, so $E_k=\pi(2n_k+1)$ for integer $n_k$. 
It is straigthforward to extend the above proof for more general transfers, such as with
long-range interactions \cite{kay2006perfect}, or for perfect fractional revivals
\cite{banchi2015perfect,genest2016quantum}. 

\subsection{Proof that CNOT needs 2-qubit terms}
\label{subsec:cnot_proof}
We here show how to use our framework to prove that a CNOT gate cannot be implemented using only one-qubit interactions.
While this result is trivial, it is nonetheless interesting to show how the framework can be used to obtain this kind of impossibility results.

The spectral decomposition of the CNOT reads
\begin{equation}
	\operatorname{CNOT} = Z_1^+ + Z_1^- X_2^+ - Z_1^- X_2^-,
	\label{eq:CNOT_definition}
\end{equation}
where we made a canonical choice for the basis of the three-fold degenerate eigenspace corresponding to the eigenvalue $+1$,
and defined $Z_i^\pm\equiv(1\pm Z_i)/2$, and similarly for $X_i^\pm$.
More explicitly, we are considering the following basis set of trace-1 projectors for the degenerate space: $\{Z_1^+ Z_2^+, Z_1^+ Z_2^-, Z_1^- X_2^+ \}$, whereas the fourth projector is bound to be $Z_1^- X_2^-$.
The corresponding principal Hamiltonian $\mathcal H_{\on{CNOT}}$, obtained by directly taking the logarithm of~\cref{eq:CNOT_definition}, is:
\begin{equation}
	\HH_{\on{CNOT}} =
	\pi Z_1^- X_2^-.
\end{equation}
Let us now consider what happens when the multivaluedness of the logarithm is taken into account, but no rotation of the degenerate eigenspace is performed.
Considering only the factors with two-qubit interactions, the following expression is found:
\begin{equation}
\begin{aligned}
	\HH_{\on{CNOT}} / 2\pi \sim \nu_{12} Z_1 Z_2 + 2\pi ( 1/2 + \nu_{43} ) Z_1 X_2,
\end{aligned}
\end{equation}
where here $\nu_{ij}\equiv \nu_i - \nu_j$, and $\nu_i\in\mathbb Z$ is the integer produced by application of the logarithm to the $i$-th projector.
Note how the $Z_1 X_2$ factor cannot be removed by \emph{any} choice of $\nu_i$,
which could be interpreted as a proof that two-qubit interaction terms are indeed necessary to implement the $\text{CNOT}$ gate.
This, however, does not in principle preclude the possibility that an appropriate rotation of the degenerate space allows to obtain a generator with only local terms.
To verify that this is not the case, we would have to consider a generic rotation $R$ of the degenerate space,
that is, an operator of the form
$R = \sum_{i,j=1}^3 r_{ij} \ketbra{+1_i}{+1_j}$,
with $\ket{+1_i}$ the $i$-th eigenvector in a fixed base of the degenerate space.
The problem is then that of finding a unitary $R$ and integers $\nu_i$ such that
\begin{equation*}
\begin{gathered}
	\nu_1 R (Z_1^+ Z_2^+) R^\dagger +
	\nu_2 R (Z_1^+ Z_2^-) R^\dagger +
	\nu_3 R (Z_1^- X_2^+) R^\dagger \\
	+ (\nu_4 + 1/2) Z_1^- X_2^-
\end{gathered}
\end{equation*}
does not contain 2-qubit interactions.
The solution of this problem is non-trivial,
mostly due to the many (9 in this case) parameters characterising a general unitary $R$.
To avoid searching solutions for such a system, we try a different approach to the problem.
Let us denote with $\tilde{\mathcal H}$ a generator with the required properties:
one that generates the same unitary as $\mathcal H_{\on{CNOT}}$ and contains only 1-qubit interaction terms.
Its general form will be:
\begin{equation}
	\tilde{\mathcal H} = h_0 +
		\sum_{\alpha=1}^3 (h_1^\alpha \sigma_1^\alpha + h_2^\alpha \sigma_2^\alpha ).
\end{equation}
As shown in the main text, for $\HTilde$ to correctly generate CNOT, it must commute with the principal generator $\HH_{\on{CNOT}}$.
Imposing this commutativity removes most of the parameters $h_i^\alpha$, leaving us with the following simplified expression:
\begin{equation}
	\tilde{\HH} = h_0 + h_{1,3} Z_1 + h_{2, 1} X_2.
	\label{eq:cnot_Hprime_reduced}
\end{equation}
The only missing step is now to impose the eigenvalues of $\HH_{\on{CNOT}} - \tilde\HH$ to be integer multiples of $2\pi$.
This gives the following system of equations:
\begin{equation}
\begin{aligned}
	&2\pi \nu_1 = (-\pi + 4 h_0 - 4 h_{1,3} - 4 h_{2,1}) / 4, \\
	&2\pi \nu_2 = (+\pi + 4 h_0 + 4 h_{1,3} - 4 h_{2,1}) / 4, \\
	&2\pi \nu_3 = (+\pi + 4 h_0 - 4 h_{1,3} + 4 h_{2,1}) / 4, \\
	&2\pi \nu_4 = (-\pi + 4 h_0 + 4 h_{1,3} + 4 h_{2,1}) / 4,
\end{aligned}
\end{equation}
with $\nu_i \in \mathbb Z$.
The above system can be seen to have no solution for $h_0, h_{1,3}, h_{2, 1}$,
therefore definitively proving that there is no rotation of the degenerate space,
and integer parameters $\nu_i$, that allow to generate the CNOT gate using only one-qubit interactions.

\subsection{Toffoli gate: derivation through conditions}
\label{subsec:toffoli_main}

We show here the details of how, using Eqs. (1), (2) and (3) in the main text, we can obtain a family of Hamiltonian generators for the Toffoli gate, containing only single- and two-qubit interactions.

The Toffoli gate can be written as
\begin{equation}
	\Toff{} = Z_1^+ + Z_1^-[Z_2^+ + Z_2^- (X_3^+ - X_3^-)],
\end{equation}
where we defined $Z_i^\pm\equiv(1\pm Z_i)/2$, and similarly for $X_i^\pm$.
The principal generator of $\Toff{}$ is $\HToff{} = \pi Z_1^- Z_2^- X_3^-$,
that is, highlighting the three-qubit interaction term,
\begin{equation}
\begin{aligned}
	\HToff =
	(\text{1- and 2-qubit terms}) - \pi/8 \, Z_1 Z_2 X_3.
\end{aligned}
\end{equation}
We start by writing down the general parametrisation of an Hamiltonian containing only one- and two-qubit interactions:
\begin{equation}
	\HTildeToff{} = h_0 \mathds1 +
	\sum h_{i,\alpha} \sigma_i^\alpha +
	\sum J_{i,j}^{\alpha,\beta} \sigma_i^\alpha \sigma_j^\beta.
\end{equation}
Assuming for simplicity that $\HTildeToff{}$ does not contain $Y_i$ components, and imposing the commutativity with $\HToff{}$, we get the following expression:
\begin{equation}
\begin{aligned}
	\HTildeToff{} =& h_0\mathds 1 +
		h_3^x X_3 + h_1^z Z_1 + h_2^z Z_2 + \\
		&J_{13}^{zx} Z_1 X_3 + J_{23}^{zx} Z_2 X_3 + J_{12}^{zz} Z_1 Z_2 + \\
		&(J_{13}^{xx} X_1 + J_{23}^{xx} X_2)(1 + X_3) + \\
		&(J_{12}^{zx} X_2 + J_{13}^{zz} Z_3)(1 + Z_1) + \\
		&(J_{12}^{xz} X_1 + J_{23}^{zz} Z_3)(1 + Z_2).
\end{aligned}
\label{eq:HTildeToff_initial}
\end{equation}
As can be directly verified, the above satisfies $[\HTildeToff{}, \HToff{}] = 0$
while also containing only one- and two-qubit interactions, and no Pauli $Y$ matrices.
We could now directly try and find the set of parameters making the eigenvalues of $\HTildeToff - \HToff$ all integer multiples of $2\pi i$, but the associated calculations are made hard by the many parameters involved.
It turns out however that we can make a number of further assumptions on the form of $\HTildeToff$, and still obtain a viable family of solutions.
One such set of assumptions, that leads to a satisfying family of solutions, is
$J_{12}^{xz} = J_{12}^{zx} = 0$,
$J_{13}^{zx} = J_{23}^{zx} = \pi/8$,
$J_{13}^{xx} = J_{23}^{xx}$,
$J_{23}^{zz} = -J_{13}^{zz}$, and
$h_1^z = h_2^z = -\pi / 8$.
With these assumptions~\cref{eq:HTildeToff_initial} becomes
\begin{equation}
\begin{aligned}
	\HTildeToff{} =
		&h_0 \mathds1 + h_3^x X_3 - \pi/8 (Z_1 + Z_2)(1 - X_3) + \\
		&J_{12}^{zz} Z_1 Z_2 + J_{13}^{xx} (X_1 + X_2)(1 + X_3) + \\
		&J_{13}^{zz} (Z_1 - Z_2)Z_3.
\end{aligned}
\end{equation}
Finally, we impose that the generator is diagonal on the first two qubits, that is, $J_{13}^{xx} = 0$.
Using this simplified expression, $\HPrimeToff{}=\HTildeToff{}-\HToff{}$ becomes
\begin{equation}
\begin{aligned}
	\HPrimeToff{} =
		&\pi/8\, Z_1 Z_2 X_3 + (h_0 - \pi/8)\mathds1 +\\
		&(h_3^x + \pi/8) X_3 + (J_{12}^{zz} - \pi/8) Z_1 Z_2 +\\
		&J_{13}^{zz} (Z_1 - Z_2)Z_3.
\end{aligned}
\label{SM:eq:toffoli_HPrime_reduced}
\end{equation}
With the above simplified expression it is now possible to directly solve the eigenvalue problem.
This results in the following family of solutions:
\begin{equation}
\begin{aligned}
	&\tilde{\mathcal H}_{\on{Toff}} =
	\frac{\pi}{8} \bigg[
	1 + 4 \left(\nu_1 + \nu_2 + 2\nu_3 + \sqrt{(\nu_3-\nu_4)^2} \right) \\
	&- (Z_1 + Z_2)(1 - X_3) + X_3 (-2 - 8\nu_1 + 8\nu_2) \\
	&+ 4 Z_1 Z_2 \left(1/4 + \nu_1 + \nu_2 - 2\nu_3 - \sqrt{(\nu_3-\nu_4)^2} \right) \\
	&+ (Z_2 - Z_1) Z_3 \,\,\sqrt{c(\nu_1, \nu_2, \nu_3, \nu_4)}
	\bigg],
\end{aligned}
\label{eq:toff_tilde_general_solution}
\end{equation}
with
\begin{equation}
\begin{split}
	c(\nu_1, \nu_2, \nu_3, \nu_4) =
		&-(1 + 4\nu_1 - 4\nu_2 + 4\nu_3 - 4\nu_4)\\
		&\times 	(1 + 4\nu_1 - 4\nu_2 - 4\nu_3 + 4\nu_4) =\\
		&\!\!= -[(1+4\nu_{12})^2-(4\nu_{34})^2],
\end{split}
\end{equation}
for all integer values of $\nu_i$ such that $c(\nu_1, \nu_2, \nu_3, \nu_4) \ge 0$.
The corresponding spectrum of $\HPrimeToff = \HTildeToff - \HToff$ is
\begin{equation}
\begin{aligned}
	\lambda_1 &= \lambda_2 = 2\pi \nu_1, \\
	\lambda_3 &= \lambda_4 = 2\pi \nu_2, \\
	\lambda_5 &= \lambda_6 = 2\pi \nu_3, \\
	\lambda_7 &= \lambda_8 = 2\pi (\nu_3 + \lvert\nu_3 - \nu_4\rvert), 
\end{aligned}
\end{equation}
while the spectrum of $\HTildeToff$ changes only in that
$\lambda_2 = 2\pi(\nu_1 + 1/2)$.
Consistently with this, $\lambda_2$ is also the eigenvalue corresponding to the non-degenerate eigenspace of $\HToff$, while all the other eigenvalues correspond to eigenvectors orthogonal to this one.
More specifically, we have
\begin{equation}
\begin{aligned}
	\ket{\lambda_1} &= \ket{0, 0, -}, \\
	\ket{\lambda_2} &= \ket{1, 1, -}, \\
	\ket{\lambda_3} &= \ket{1, 1, +}, \\
	\ket{\lambda_4} &= \ket{0, 0, +}, \\
	\ket{\lambda_5} &= \ket{1, 0} \otimes N_5\left[(a - b) \ket0 + \ket1 \right], \\
	\ket{\lambda_6} &= \ket{0, 1} \otimes N_6\left[(a + b) \ket0 + \ket1 \right], \\
	\ket{\lambda_7} &= \ket{1, 0} \otimes N_6\left[(a + b) \ket0 - \ket1 \right], \\
	\ket{\lambda_8} &= \ket{0, 1} \otimes N_5\left[(a - b) \ket0 - \ket1 \right],
\end{aligned}
\label{eq:toffoli_eigenvectors_solution}
\end{equation}
where
\begin{equation}
\newcommand{\denom}{1 + \bar\nu_{12}}
\begin{gathered}
	a = \frac{\lvert\bar\nu_{34}\rvert}{\denom{}},
	\qquad
	b = \frac{\sqrt{\bar\nu_{34}^2 - (\bar\nu_{12} + 1)^2}}{\denom{}},
\end{gathered}
\end{equation}
It is worth noting that the orthogonality of these eigenvectors follows from the easily verified property of the above coefficients: $a^2 - b^2 = 1$.
Furthermore, we note that $c(\nu_1, \nu_2, \nu_3, \nu_4) \ge 0$ cannot be satisfied unless $\nu_3 \neq \nu_4$.
This in turn, looking at \cref{eq:toffoli_eigenvectors_solution},
reveals that all the solutions are made possible by a non-trivial lifting of the degeneracy of the subspaces $\ketbra{0,1}$ and $\ketbra{1,0}$.
Let us now try to understand how and why the derived $\HPrimeToff{}$ works.
Let us use the notation $P_i \equiv \ketbra{\lambda_i}$,
and consider the projector over the last two eigenvectors.
Highlighting the 3-qubit terms, we find
\begin{equation}
\begin{aligned}
	P_7 \simeq - N_6^2 \frac{Z_1 Z_2}{4} \bigg[
		\big((a + b)^2 - 1\big) \frac{Z_3}{2}
		- (a + b) X_3
	\bigg], \\
	P_8 \simeq - N_5^2 \frac{Z_1 Z_2}{4} \bigg[
		\big((a - b)^2 - 1\big) \frac{Z_3}{2}
		- (a - b) X_3
	\bigg].
\end{aligned}
\end{equation}
The term in the Hamiltonian to which these two projectors contribute is
$2\pi \nu_{3,4} (P_7 + P_8)$, with $\nu_{3,4} = \nu_3 + \lvert \nu_3 - \nu_4\rvert$.
A little algebra reveals that the 3-qubit terms in $P_7 + P_8$ are
\begin{equation}
\begin{aligned}
	- \frac{Z_1 Z_2 Z_3}{8} \left[
		N_6^2\big((a+b)^2 - 1\big)
		+ N_5^2 \big((a-b)^2 - 1\big)
	\right] \\
	+ \frac{Z_1 Z_2 X_3}{4} \left[
		N_6^2(a+b) + N_5^2(a-b)
	\right].
\end{aligned}
\end{equation}
Recalling the definitions of $a,b,N_5,N_6$, we see that the coefficient of $Z_1 Z_2 Z_3$ vanishes, and the resulting expression becomes
\begin{equation}
	P_7 + P_8 = (...) +
	Z_1 Z_2 X_3 \frac{1 + 4(\nu_1 - \nu_2)}{16\lvert \nu_3 - \nu_4\rvert}.
\end{equation}
Substitution of the appropriate values of $\nu_i$ shows that the above term can be used to generate the 3-qubit factor $\pi/8 \,\,Z_1 Z_2 X_3$,
\emph{without introducing additional 3-qubit factors}.
In Box~\ref{tcolorbox:toffoli} are given the full expressions for the projectors and the found solutions for the Toffoli gate.
It is also interesting to note that all of the above still holds if the $X_i$ operators are replaced with $Y_i$ operators.
That is, the expressions found solving for the Toffoli, by simple substitution $X_i \to Y_i$,
also give a generator with only 2-qubit interactions for the CCY gate
(that is, the gate that applies $Y$ to the third qubit conditionally to the first 2 qubits being in the $\ket1$ state).

\begin{tbox}[label=tcolorbox:toffoli]{Toffoli}
	\begin{equation*}
	\begin{aligned}
		P_1 &= Z_1^+ Z_2^+ X_3^-,
		\qquad
		P_2 &= Z_1^- Z_2^- X_3^-,
		\qquad
		P_3 &= Z_1^+ Z_2^+ X_3^+,
		\qquad
		P_4 &= Z_1^- Z_2^- X_3^+,
	\end{aligned}
	\end{equation*}
	\begin{equation*}
	\begin{aligned}
		P_5 &= Z_1^- Z_2^+ \frac{1}{2\lvert\bar{\nu}_{34}\rvert}
		\left[
			\lvert\bar\nu_{34}\rvert +
			(1 + \bar\nu_{12}) X_3 -
			\sqrt{-(1 + \bar\nu_{12})^2 + \bar\nu_{34}^2} Z_3
		\right], \\
		P_6 &= Z_1^+ Z_2^- \frac{1}{2\lvert\bar{\nu}_{34}\rvert}
		\left[
			\lvert\bar\nu_{34}\rvert +
			(1 + \bar\nu_{12}) X_3 +
			\sqrt{-(1 + \bar\nu_{12})^2 + \bar\nu_{34}^2} Z_3
		\right], \\
		P_7 &= Z_1^- Z_2^+ \frac{1}{2\lvert\bar{\nu}_{34}\rvert}
		\left[
			\lvert\bar\nu_{34}\rvert -
			(1 + \bar\nu_{12}) X_3 +
			\sqrt{-(1 + \bar\nu_{12})^2 + \bar\nu_{34}^2} Z_3
		\right], \\
		P_8 &= Z_1^+ Z_2^- \frac{1}{2\lvert\bar{\nu}_{34}\rvert}
		\left[
			\lvert\bar\nu_{34}\rvert -
			(1 + \bar\nu_{12}) X_3 -
			\sqrt{-(1 + \bar\nu_{12})^2 + \bar\nu_{34}^2} Z_3
		\right].
	\end{aligned}
	\end{equation*}
	\begin{equation*}
		P_1 + P_2 = \frac{1}{4} (1 + Z_1 Z_2) (1 - X_3),
		\qquad
		P_3 + P_4 = \frac{1}{4} (1 + Z_1 Z_2) (1 + X_3).
	\end{equation*}
	\begin{align*}
		P_5 + P_6 = \frac{1}{4\lvert\bar\nu_{34}\rvert} \left[
			(1 - Z_1 Z_2) \lvert\bar\nu_{34}\rvert +
			(1 - Z_1 Z_2) X_3 (1 + \bar\nu_{12}) +
			(Z_1 - Z_2)Z_3 \sqrt{\bar\nu_{34}^2 - (1 + \bar\nu_{12})^2}
		\right], \\
		P_7 + P_8 = \frac{1}{4\lvert\bar\nu_{34}\rvert} \left[
			(1 - Z_1 Z_2) \lvert\bar\nu_{34}\rvert -
			(1 - Z_1 Z_2) X_3 (1 + \bar\nu_{12}) -
			(Z_1 - Z_2)Z_3 \sqrt{\bar\nu_{34}^2 - (1 + \bar\nu_{12})^2}
		\right].
	\end{align*}
	It is easily verified from the above that
	\begin{equation*}
		P_1 + P_2 + P_3 + P_4 = \frac{1}{2} (1 + Z_1 Z_2),
		\qquad
		P_5 + P_6 + P_7 + P_8 = \frac{1}{2} (1 - Z_1 Z_2),
	\end{equation*}
	so that the sum of the projectors gives the identity as it should.
	On the other hand, multiplying by the appropriate $\nu_i$ factors, we get
	\begin{equation*}
	\begin{aligned}
		2\pi\left[\nu_1(P_1 + P_2) + \nu_2(P_3 + P_4)\right]
		&= (...) + \frac{\pi}{2} (\nu_2 - \nu_1) Z_1 Z_2 X_3, \\
		2\pi\left[\nu_3(P_5 + P_6) + \nu_4(P_7 + P_8)\right]
		&= (...) + \frac{\pi}{2} (\nu_1 - \nu_2) Z_1 Z_2 X_3 + \frac{\pi}{8} Z_1 Z_2 X_3,
	\end{aligned}
	\end{equation*}
	with the last identity holding for $\nu_3\neq \nu_4$.
\end{tbox}

A different way to understand $\HTildeToff$ is to analyse the four two-dimensional subspaces on the main diagonal, exploiting the fact that $\HTildeToff{}$ acts diagonally on the first two qubits.
Straightforward calculations lead to
\begin{equation*}
\begin{aligned}
	\mel{00}{\HTildeToff}{00} &= \pi \left[(\nu_1+\nu_2)-(\nu_1-\nu_2)X\right], \\
	\mel{01}{\HTildeToff}{01} &= 2\pi\nu_3 + \pi\lvert\nu_{34}\rvert(1 - \sigma_{01}), \\
	\mel{10}{\HTildeToff}{10} &= 2\pi\nu_3 + \pi\lvert\nu_{34}\rvert(1 - \sigma_{10}), \\
	\mel{11}{\HTildeToff}{11} &= \frac{\pi}{2}\left[
		(1+2(\nu_1+\nu_2))-(1+2\nu_{12})X
	\right], \\
\end{aligned}
\end{equation*}
where
\begin{equation}
\begin{aligned}
	\sigma_{01}&\equiv\frac{(1+4\nu_{12})X + \sqrt{c}Z}{4\lvert\nu_{34}\rvert}, \\
	\sigma_{10}&\equiv\frac{(1+4\nu_{12})X - \sqrt{c}Z}{4\lvert\nu_{34}\rvert}.
\end{aligned}
\end{equation}
It can be verified that for all values of $\nu_1, \nu_2, \nu_3, \nu_4$, the two-dimensional identity and $X$ are correctly generated in the $\ket{00}$ and $\ket{11}$ spaces, respectively.
On the other hand, in the $\ket{01}$ and $\ket{10}$ spaces, the two-dimensional identity is generated as long as $\nu_3\neq\nu_4$, as was also derived before.

In particular, the class of solutions given by $\nu_1 = \nu_2 = \nu_3 = 0$ is
\begin{equation}
\begin{split}
	\frac{\pi}{8} \bigg[
	1 + 4\lvert\nu_4\rvert
	- 2 X_3 - Z_1 - Z_2
	+ (Z_1 + Z_2) X_3 \\
	+ Z_1 Z_2 (1 - 4\lvert\nu_4 \rvert)
	+ (Z_2 - Z_1) Z_3 \sqrt{16\nu_4^2 - 1}
	\bigg],
\end{split}
\label{SM:eq:toffoli_generator_nu4}
\end{equation}
for all $\nu_4\neq 0$.
It is interesting to look at the explicit form of the exponentials
generated by this class generators.
Computing $\exp(i \tilde{\mathcal H} t)$ with $\tilde{\mathcal H}$ as in \cref{SM:eq:toffoli_generator_nu4}, we get the following unitary:
\begin{equation}
	\begin{pmatrix}
		\mathds1_2 & \mathbb0 & \mathbb0 & \mathbb0 \\
		\mathbb0 & S(t, \nu_4) & \mathbb0 & \mathbb0 \\
		\mathbb0 & \mathbb0 & S(t, \nu_4) & \mathbb0 \\
		\mathbb0 & \mathbb0 & \mathbb0 & X(t)
	\end{pmatrix},
\end{equation}
where
\begin{equation}
	S(t, \nu_4) = \begin{pmatrix}
		a + b & c \\
		c & a - b
	\end{pmatrix},
\end{equation}
\begin{equation}
\begin{gathered}
	a = \frac{1 + e^{2i\pi t \nu_4}}{2},
	\qquad c = \frac{1 - e^{2i\pi t \nu_4}}{8\nu_4},  \\
	b = \frac{(-1 + e^{2i\pi t \nu_4})\sqrt{16\nu_4^2 - 1}}{8\nu_4},
\end{gathered}
\end{equation}
and
\begin{equation}
	X(t) = \frac{1}{2} \begin{pmatrix}
		(1 + e^{i\pi t}) & (1 - e^{i\pi t}) \\
		(1 - e^{i\pi t}) & (1 + e^{i\pi t})
	\end{pmatrix}
\end{equation}
For large (in modulus) values of $\nu_4$,
$a + b \to e^{2i\pi t \nu_4}$, $a - b \to 1$ and $c\to0$,
so that the exponential becomes
\begin{equation}
	\begin{pmatrix}
		\mathds1 & & & & &\\
		& e^{2i\pi t \nu_4} & & & & \\
		& & 1 & & & \\
		& & & e^{2i\pi t \nu_4} & & \\
		& & & & 1 & \\
		& & & & & X(t)
	\end{pmatrix},
\end{equation}
which very closely resembles the matrix obtained by exponentiating the principal generator
$\mathcal H_{\on{Toff}} = \pi Z_1^- Z_2^- X_3^-$:
\begin{equation}
	\exp(i t \mathcal H_{\on{Toff}}) =
	\begin{pmatrix}
		\mathds1 & & & \\
		& \mathds1 & & \\
		& & \mathds1 & \\
		& & & X(t)
	\end{pmatrix}.
\end{equation}
A different solution derived from~\cref{eq:toff_tilde_general_solution} is
\begin{equation}
\begin{aligned}
	\tilde{\mathcal H}_{\on{Toff}} =
	\frac{9\pi}{8} + \frac{3\pi}{4} X_3 - \frac{\pi}{8} (Z_1 + Z_2)
	+ \frac{\pi}{8} Z_1 Z_2 \\
	+ \frac{\pi}{8} (Z_1 + Z_2) X_3
	- \frac{\sqrt7 \pi}{8} (Z_1 - Z_2) Z_3.
\end{aligned}
\end{equation}
Moreover, it is worth noting that~\cref{eq:toff_tilde_general_solution} is only one possible family of solutions, and that different assumptions will lead to different ones.
For example, a similar reasoning as above, starting however from the assumptions $J_{23}^{zz}=J_{13}^{zz}$ will lead to solutions such as (note the use of $(Z_1+Z_2)$ terms here, making this solution not derivable from~\cref{eq:toff_tilde_general_solution}):
\begin{equation}
\begin{aligned}
	\HTildeToff{} =\,
		&\frac{9\pi}{8} - \frac{7\pi}{8}X_3 + \frac{\sqrt{15}\pi}{8} Z_3 + \frac{\pi}{8}Z_1 Z_2 \\
		&\frac{\pi}{8}(Z_1+Z_2)\left(-1+\frac{5}{2}X_3 + \frac{\sqrt{15}}{2}Z_3\right).
\end{aligned}
\end{equation}

\subsection{Toffoli gate: an example of direct \emph{a posteriori} derivation}
\label{subsec:toffoli_posteriori_derivation}
We will here show a line of thought that could have conceivably led to \cref{SM:eq:toffoli_generator_nu4} (in the case $\nu_4=1$), by direct analysis, and without using any of the tools shown in the paper.
It will be useful to keep in mind the expressions of $Z_1 \pm Z_2$:
\begin{equation}
\begin{aligned}
	Z_2 + Z_1 &= \on{diag}(2, 2, 0, 0, 0, 0, -2, -2), \\
	Z_2 - Z_1 &= \on{diag}(0, 0, -2, -2, 2, 2, 0, 0).
\end{aligned}
\end{equation}
Given that we want to generate a CC-X gate, and remembering that
$\exp\left[\frac{i\pi}{2}(1-X)\right] = X$, it is reasonable to start building our Hamiltonian as
\begin{equation}
	\mathcal H_1 = - \pi \left(\frac{Z_1+Z_2}{2}\right)\left(\frac{1-X_3}{2}\right),
\end{equation}
which however will generate an $X$ evolution both in the $\ket{00}$ and in the $\ket{11}$ sectors, while we want it only in the latter sector:
	$\mathcal H_1 \doteq \on{diag}(-X^-, 0, 0, X^-)$.
We can remove the term in the $\ket{00}$ sector exploiting the sign difference introduced by $Z_1 + Z_2$, by directly adding an appropriate 1-qubit interaction term:
\begin{equation}
\begin{aligned}
	\mathcal H_2 &= \frac{1}{2} \left[ \mathcal H_1 + \frac{\pi}{2} (1 - X_3) \right] \\
				 &= \pi \on{diag}(0, X^-\!/2, X^-\!/2, X^-),
\end{aligned}
\label{eq:toffoli_analytical_step2}
\end{equation}
where we remember that $\exp(i\pi X^-) = X$.
\Cref{eq:toffoli_analytical_step2} now correctly reproduces the evolution on $\ket{00}$ and $\ket{11}$, but also wrongly evolves $\ket{01}$ and $\ket{10}$.
To remove these additional terms while at the same time leaving the others unaffected we use the fact that
$\exp(i\pi (1 \pm \bs{\sigma})) = \mathds1$, for any normalised vector of sigma matrices: $\bs\sigma \equiv \sum_{i=1}^3 n_i \sigma_i$ with $n_1^2 + n_2^2 + n_3^2 = 1$.
To convert the central terms in~\cref{eq:toffoli_analytical_step2} into something like this we observe that we can rewrite the second term in the above equation as
\begin{equation}
	\pi/4(1-X_3) = \pi/8(2 - 2 X_3) = \pi/8(5 - 3 - 2 X_3).
\end{equation}
Remembering that $Z_1 Z_2 = \on{diag}(1, -1, -1, 1)$, we substitute the above with
$\pi/8(5 - 3Z_1 Z_2 - 2X_3)$.
This change affects only the central terms, converting the expression into:
$\pi \on{diag}(0, 1 - X/4, 1 - X/4, X^-)$.
The reason this form is preferable is that we can now simply add a factor in the central terms to convert them into an expression of the form $1 - \bs\sigma$.
Adding an interaction of the form $\pi \alpha (Z_2 - Z_1)Z_3$ gives
\begin{equation}
	\pi\on{diag}(0, 1 - X/4 - 2\alpha Z, 1 - X/4 + 2\alpha Z, X^-).
\end{equation}
For the central terms to exponentiate to the identity we need them to become of the form $1-\bs\sigma$ with normalised $\bs\sigma$.
This is easily achieved by choosing $\alpha=\pm\sqrt{15}/8$.
The final expression is therefore:
\begin{equation}
\begin{aligned}
	8/\pi \,\, \mathcal H_3 = -(Z_1 + Z_2)(1 - X_3) \\+ (5 - 3Z_1Z_2 - 2 X_3)
				 \pm \sqrt{15} (Z_2 - Z_1)Z_3.
\end{aligned}
\end{equation}
Note that instead of $\pi \alpha (Z_2 - Z_1)Z_3$ we could have equivalently chosen
$\pi\alpha(Z_2-Z_1)O_3$ for any $O_3 = a Y_3 + b Z_3$ and $a^2 + b^2 = 1$.
The above reasoning explains the origin of the weird $\sqrt{15}$ factor: it comes as the coordinate necessary to make the vector unitary: for $X/4 + x Z / 4$ to be normalized, $x = \sqrt{15}$ must be satisfied.

\section{Supervised learning approach}
\label{sec:supervised_learning_approach}
We here study more in depth the following problem: given a target gate $\G$ and a parametrised Hamiltonian ${\mathcal H(\bs\lambda) = \sum_k \lambda_k \sigma_k}$, with $\lambda_k\in\mathbb R$ and $\sigma_k$ Hermitian operators, what is the set of parameters $\bs\lambda_0$ such that $\exponential(i\HH(\bs\lambda_0)) = \G$?
We present a supervised learning approach to numerically solve this problem, considerably extending the ideas presented in Ref.~\cite{banchi2016quantum}.
Thanks to a number of numerical optimisation techniques, and in particular the use of \ac{AD}~\cite{baydin2015automatic,bartholomewbiggs2000automatic,wengert1964a,bischof2008advances}, we can explore a variety of different scenarios, optimising over potentially hundreds of Hamiltonian parameters.
On top of this, condition (1b) in the main text is used to further speed-up the numerical training, removing many interaction parameters that are known not to lead to the target gate.

\subsection{Supervised learning}
\label{subsec:supervised_learning}

Supervised learning is the task of inferring or approximating a function, given a set of pre-labeled data~\cite{bishop2006pattern,mohri2012foundations}.
A supervised learning algorithm starts with some \emph{model} --- a functional relation $g_{\bs\lambda}$ parametrised by a set of parameters $\bs\lambda$ --- and finds a $\bs\lambda_0$ making $g_{\bs\lambda_0}$ as close as possible to a target function $f$.
To do this, a set of pre-labeled \emph{training data} $\{ (x_1, y_1), (x_2,y_2), ...\}$ is used,
where here $y_k=f(x_k)$ is the output that we want the algorithm to associate to the input $x_k$.

Among the most used supervised learning models are \acp{NN}~\cite{hechtnielsen1989theory,haykin1998neural}.
These are parametric non-linear models which play a prominent role in many machine learning tasks, such as dimensionality reduction, classification, and feature extraction~\cite{hechtnielsen1989theory,haykin1998neural}.
\acp{NN} have also recently proven useful for several problems in quantum many-body theory~\cite{amin2016quantum,wang2016discovering,hush2017machine,carleo2017solving,carrasquilla2017machine,torlai2017manybody,broecker2017quantum,deng2017quantum},
quantum compilation~\cite{swaddle2017generating}, quantum stabilizer codes~\cite{krastanov2017deep} and entanglement quantification~\cite{gray2017measuring}.

A \ac{NN} is \emph{trained} by optimising its parameters using a dataset of pre-labelled data.
A common  way to do this is use variations of a gradient-descent-based technique named \ac{SGD}.
Gradient descent algorithms aim to optimize a problem function $f(\bs x)$, starting from an initial point $\bs x_0$ and performing a number of small steps towards the direction of maximum slope (that is, $\grad f(\bs x)$).
The optimal point $\bsx_{\on{opt}}$ is thus obtained via a sequence of small perturbations of the point $\bsx$, which starting from $\bsx_0$ reaches the nearest local stationary point by following the slope.
In the simplest version of the algorithm, the update rule is simply $\bs x \to \bs x - \eta \grad f(\bs x)$, with $\eta$ a small real parameter commonly referred to as \emph{learning rate}.
\ac{SGD}, on the other hand, is suitable for a situation in which one is given a parametrised functional relationship of the form $f(\bs x; \bs w)$, and asked for a set of ``parameters'' $\bs w_0$ such that $f(\bs x; \bs w_0)$ is minimum (maximum) for all \emph{inputs} $\bs x$.
Such a case can be handled via \ac{SGD}, which in its simplest form involves picking a random $\bs x_1$, executing a number of gradient descent iterations over $\bs w$, then picking a new $\bs x_2$ and iterating the procedure.
The updating rule for \ac{SGD} is therefore of the form
\begin{equation}
	\bs w \to \bs w - \eta\grad_{\bs w}f(\bs x;\bs w).
	\label{eq:updating_rule}
\end{equation}
While standard gradient descent, being a local optimisation algorithm, is liable to getting stuck in local minima, \ac{SGD} can at least partially avoid this issue, in that generally a local minimum for an input $\bs x$ is not a local minimum for a different input $\bs x'$.
Many variations of \ac{SGD} are used in different circumstances.
For example, in the so-called \emph{mini-batch} \ac{SGD}, instead of updating with a single input $\bs x$, one uses a \emph{batch} of inputs $\{\bs x_1,...,\bs x_M\}$, and updates the parameters using the averaged gradient:
$\bs w\to\bs w-\eta\sum_{k=1}^M\grad_{\bs w}f(\bs x_k; \bs w)/M$.
More sophisticated updating rules are used to increase the training efficiency in different circumstances.
Common techniques involve dynamically updating the learning rate, or using \emph{momentum gradient descent}~\cite{goh2017momentum,ruder2016overview} techniques.

To see how this class of optimisation problems is relevant to us, consider the fidelity function $\F$ defined as
\begin{equation}
	\F(\bs\lambda, \psi) \equiv \mel{\psi}{\G^\dagger \exp(i \HH(\bs\lambda)) \G}{\psi},
	\label{eq:def_fidelity}
\end{equation}
with $\G$ the target gate, $\bs\lambda$ the set of parameters, and $\psi$ an input state.
The gate design problem is then equivalent to finding $\bs\lambda$ such that $\F(\bs\lambda, \psi)$ is maximised (that is, equal to 1) for all $\psi$.
One possibility to solve this problem is to consider the average fidelity $\bar{\F}(\bs\lambda)$, for which explicit formulas are known~\cite{banchi2011nonperturbative,pedersen2007fidelity,magesan2011gate}.
Standard optimisation methods, like standard gradient descent or differential evolution, can be applied directly on $\bar{\F}(\bs\lambda)$.
This, however, reveals to be impractical, due to the complexity of the associated parameter landscapes.
On the other hand, \ac{SGD} allows to use a simple and efficient local maximisation method, while at the same time being less prone to getting stuck in local maxima.
This works particularly well in this case, because we know that the sets of parameters corresponding to the target gate are \emph{all and only} those such that \emph{for all inputs $\psi$} the fidelity is unitary.

A crucial step, efficiency-wise, in gradient descent algorithms, is the evaluation of the gradient.
Numerically approximating the gradient, as done in previous works~\cite{banchi2016quantum}, is generally inefficient and scales badly with the number of optimised parameters.
Here we will instead make use of the powerful technique of \acf{AD}~\cite{bartholomewbiggs2000automatic,bischof2008advances}, described in~\cref{subsec:backpropagation}.
\ac{AD} dramatically improves the training efficiency, allowing to explore a richer variety of circumstances.


\subsection{Backpropagation}
\label{subsec:backpropagation}
The gradient evaluation phase is efficiency-wise crucial for the training of a neural network.
Computing the partial derivatives of the cost function with a standard method, like finite differences, has a complexity $\mathcal O(N_{\bs w}^3)$, with $N_{\bs w}$ the number of parameters to differentiate~\cite{bishop2006pattern}.
This inefficiency can however be avoided using \emph{error backpropagation} via \ac{AD}.
With this technique, the complexity of the gradient evaluation phase can be cut down to $\mathcal O(N_{\bs w}^2)$~\cite{bishop2006pattern}.
This works by first decomposing the cost function of the model in terms of elementary operations, that is, functions the gradient of which is known analytically.
In this way the \emph{computational graph} representing the functional relation between input and output is built.
A computational graph is a directed acyclic graph, whose nodes represent the operations, and edges the flowing direction of inputs into outputs (see~\cref{fig:automatic-differentiation}).
Once the computational graph is built, the gradients with respect to the model parameters can be computed efficiently.
This happens in two stages, as schematically illustrated in~\cref{fig:automatic-differentiation}.
At first, every node of the computational graph is progressively computed, starting from the inputs (the current values of the model parameters) up to the final value of the error function.
During this process, the intermediate values of the elementary operations are cached.
This is the so-called \emph{feed-forward} phase.
The second phase (so-called \emph{backpropagation} phase) starts from the output, and consists of progressively computing the gradients of the error function with respect to the independent variables.

To better understand \ac{AD}, let us consider a simple example.
Suppose the error function of the model is of the form
$g(\bs w)\equiv f(\bs f^{(2)}(\bs f^{(1)}(\bs w)))$,
where $\bs w$ is a set of parameters, and $\bs f^{(i)}$ are intermediate ``elementary'' functions, the gradients of which are supposed to be known analytically.
Making use of the chain rule, the gradient of $g$ reads
\begin{equation}
\begin{split}
	\grad g(\bs w) =
	\sum_{k} \partial_{k} f(\bs y^{(2)})
	\grad f_{k}^{(2)}(\bs y^{(1)}),
\end{split}
\label{eq:grad_nested_f}
\end{equation}
where $\bs y^{(2)} = \bs f^{(2)}(\bs f^{(1)}(\bs w))$ and
$\bs y^{(1)} = \bs f^{(1)}(\bs w)$.
During the feed-forward phase the values of $\bs y^{(1)}$ and then $\bs y^{(2)}$ are progressively computed and cached.
Using $\bs y^{(2)}$, and the known expression for $\partial_k f$, $\partial_{k} f(\bs y^{(2)})$ is then efficiently computed.
The process continues by evaluating $\grad f_k^{(2)}$, which is written as
\begin{equation}
	\grad f_k^{(2)}(\bs y^{(1)}) =
	\sum_j \partial_j f_k^{(2)}(\bs y^{(1)}) \grad f_j^{(1)}(\bs w).
\end{equation}
Again, being $\bs y^{(1)}$ already computed during the feed-forward, $\partial_j f_k^{(2)}(\bs y^{(1)})$ is readily computed.
The last component needed for the full gradient is $\partial_i f_j^{(1)}(\bs w)$, all parts of which are known.
This method therefore allows to efficiently evaluate numerical the gradient of complicated functions, without approximating the derivatives.

In the context of training neural networks, the function to be derived is the \emph{cost function} of the network, that is, roughly speaking, the (euclidean) distance between the result obtained for an input and the corresponding training output.
For the gate design problem, we will use another notion of \emph{distance} between output obtained and output expected.
For quantum states, the fidelity between these turns out to work well.

\begin{figure*}[tb]
	\centering
	\begin{minipage}{0.3\linewidth}
		\centering
		\includegraphics[height=1.3\textwidth]{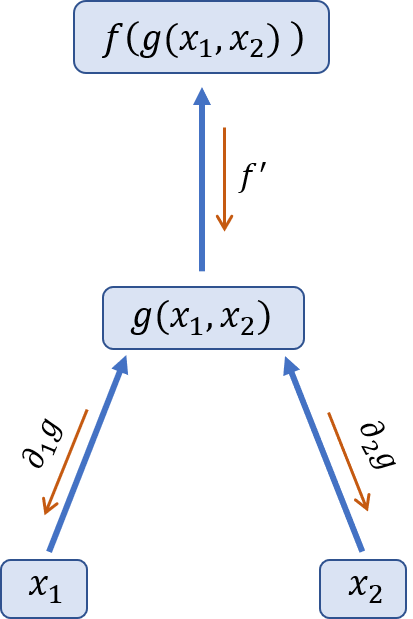}
	\end{minipage}\hfill
	\begin{minipage}{0.3\linewidth}
		\centering
		\includegraphics[height=1.3\textwidth]{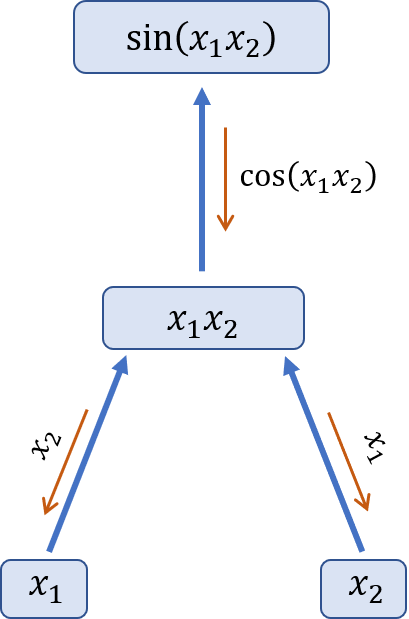}
	\end{minipage}\hfill
	\begin{minipage}{0.3\linewidth}
		\centering
		\includegraphics[height=1.3\textwidth]{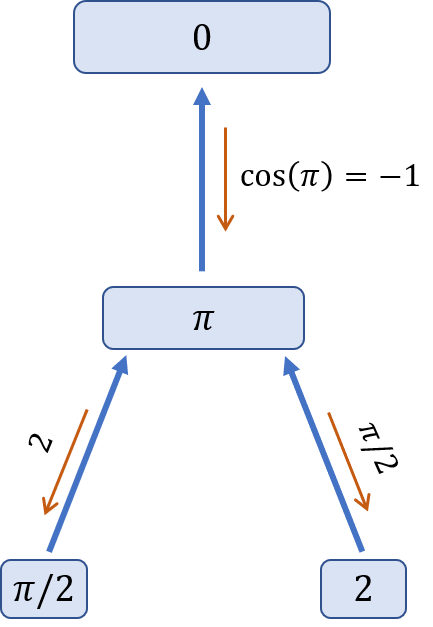}
	\end{minipage}
	\caption{
		Examples of \ac{AD} in backpropagation mode.
		\textbf{(a)}
		Schematic representation of \ac{AD} of a function with one output and two inputs.
		Starting from numerical values for $x_1$ and $x_2$, one computes $g(x_1, x_2)$ and then $f(g(x_1, x_2))$.
		To get $\grad f(g(x_1, x_2))$, one then computes $f'(g(x_1, x_2))\partial_i g(x_1, x_2)$.
		Note that all components of this expression are known: $f'$ and $\partial_i g$ are known by assumption, and the value of $g(x_1, x_2)$ has been computed and cached in the forward propagation phase.
		\textbf{(b)} Example of application of \ac{AD} to compute the gradient of $\cos(x_1 x_2)$.
		\textbf{(c)} Using the same example function as (b), we give an example of the actual number computed at all stages when the inputs are $(x_1, x_2) = (\pi / 2, 2)$.
	}
	\label{fig:automatic-differentiation}
\end{figure*}

\subsection{Implementation details}
\label{subsec:implementation_details}

We used python as language of choice for the implementation of the supervised learning.
Being Python language of widespread use in the machine learning community, many libraries and frameworks are available to build computational graphs over which \ac{AD} can be used.
In particular, we used \textsc{Theano}~\cite{team2016theano}, together with the \textsc{QuTiP} library for the simulation of the dynamics of quantum systems~\cite{johansson2012qutip,johansson2013qutip}.

Our implementation allows the training of an arbitrary target gate, parametrised via a time-independent Hamiltonian $\mathcal H(\bs\lambda)$.
The parametrisation is completely arbitrary (provided the dependence on the parameters is linear), so that the Hamiltonian can be chosen as $\mathcal H(\bs\lambda)=\sum_i \lambda_i A_i$ for any set of matrices $A_i$ and number of parameters $\lambda_i$.
This is made possible by the flexibility of \ac{AD}, which allows to automatically build an efficiently differentiable computational graph, without needing to hardcode the structure of the Hamiltonian.

The goal of the algorithm is, given a target gate $\mathcal G$ and a parametrisation for the Hamiltonian $\mathcal H(\bs\lambda)$, find the $\bs\lambda_0$ such that $\exp(i\mathcal H(\bs\lambda_0))=\mathcal G$.
We use for the purpose mini-batch \ac{SGD} with momentum.
The \emph{mini-batch} version of \ac{SGD} involves computing the gradient, at every iteration, averaging over the gradients computed for a number of states.
Making such \emph{batches} of states larger or smaller allows to enhance or decrease the variance of the gradients with respect to the input state.
The use of \emph{momentum}~\cite{ruder2016overview,goh2017momentum} involves using a modified version of~\cref{eq:updating_rule}.
The updating rule is instead given by
\begin{equation}
\begin{aligned}
	\bs v &\to \gamma \bs v + \eta \grad_{\bs\lambda} \F(\psi, \bs\lambda), \\
	\bs\lambda &\to \bs\lambda + \bs v,
\end{aligned}
\label{eq:updating_rule_momentum}
\end{equation}
where here $\eta$ is the \emph{learning rate} and $\gamma$ the \emph{momentum}.
The use of the auxiliary parameter $\bs v$ during the training discourages sudden changes of direction, and can make the training significantly more efficient~\cite{goh2017momentum}.

While the cost function $\F$ is always real, some of the intermediate calculations needed to compute it involve complex numbers.
While this poses no fundamental problems, many of the \ac{ML} libraries do not support \ac{AD} over functions with complex inputs or outputs.
We worked around this problem using a similar trick to the one reported in~\cite{leung2017speedup}.
In particular, to use the existing framework, we mapped the problem into one involving only real numbers.
To do this, we map complex matrices into real ones via the bijection
$A\mapsto\mathfrak{Re}(A)\equiv\mathds1\otimes A_{R} - i \sigma_y\otimes A_{I}$,
where $A_R$ and $A_I$ are the real and imaginary parts of $A$, respectively.
At the same time, state vectors are to be mapped to
$\Psi\mapsto\mathfrak{Re}(\Psi)\equiv(\Psi_R, \Psi_I)^T$.
It is easy to verify that with this mapping
$A\Psi\mapsto\mathfrak{Re}(A\Psi)=\mathfrak{Re}(A)\mathfrak{Re}(\Psi)$,
so that all calculations can be equivalently be carried out with the real versions of matrices and vectors.

More specifically, the employed algorithm involves the following steps:
\begin{enumerate}
	\item Choose an initial set of parameters $\bs\lambda$ (randomly, or specific values if one has an idea of where a solution might be).
	A number of other hyperparameters have to be decided at this step, depending on the exact \ac{SGD} method used. In particular, for mini-batch \ac{SGD} with momentum and decreasing learning rate, one has to decide the momentum $\gamma$, the initial value of $\eta$, the rate at which $\eta$ decreases during the training, and the size $N_b$ of the batches of states used for every gradient descent step.
	\item Repeat the following loop $N_e$ times, or until a satisfactory result is obtained.
	Each such iteration is conventionally named an \emph{epoch}.
	Another hyperparameter to be chosen beforehand is the number of training states $N_{tr}$ to be used in each epoch.
	Once this is fixed, every epoch will involve a number $N_{tr}/N_e$ of gradient descent steps, each one using $N_e$ states for a single gradient calculation.
	$N_e$ random training states are sampled, to be used during the epoch.
	\begin{enumerate}
		\item Pick $N_b$ of the $N_e$ training states.
		\item Forward-propagate each state of the sample, and then backpropagate the gradients, thus computing the average gradient over the mini-batch $\grad_{\bs\lambda} \F(\bs\lambda)$.
		\item Update the coupling strengths $\lambda$ as per~\cref{eq:updating_rule_momentum}.
		\item Return to point (a).
	\end{enumerate}
\end{enumerate}

\subsection{Results}
\label{subsec:numerical_results}

A sample of training results for Toffoli, Fredkin, and "double Fredkin" gates are given in Fig. 1 (a), (b), and (c) in the main text.
In~\cref{fig:toffoli_diagonal_parhistories,fig:fredkin_diagonal_parhistories,fig:doublefredkin_diagonal_parhistories} are shown the training histories of the parameters for eight different solutions for Toffoli, Fredkin and \emph{double Fredkin}, respectively.
These illustrate how quickly the networks converge for different initial values of the parameters.
In all the shown cases the target gates are obtained with unit fidelity up to numerical precision (that is, all fidelities are between $1-10^{-16}$ and $1$).
Different sets of optimisation hyperparameters are found to give acceptable solutions.
For the trainings shown in this paper we used a dynamically updated learning rate given, for the $k^{\text{th}}$ epoch, by $\eta=1/(1 + k \alpha)$ with the \emph{decay rate} $\alpha=0.005$.
The other hyperparameters were chosen as
$\gamma=0.5$, 
$N_b = 2$, $N_{tr} = 200$.
Different initial values for the parameters were tested, but in most cases we started the training with either vanishing or random (following a normal distribution) parameters.
For the training of the four-qubit gate we found the network to converge sooner to a solution when the parameters were initialised to a positive value (often with all parameters initialised to $4$).

In~\cref{fig:toffoli_fidVSpars,fig:fredkin_fidVSpars,fig:doublefredkin_fidVSpars} we report the behaviour of the fidelity upon changes of the learnt Hamiltonian parameters, for Toffoli, Fredkin and \emph{double Fredkin} gates, respectively.
As shown in these plots, the stability of the implemented gates with respect to variations of time and interactions values greatly varies between different solutions, as well as between different parameters in the same solutions.

To assess the feasibility of nontrivial gates in more restrictive experimental scenarios, we performed a systematic analysis of the reachability of Fredkin and Toffoli gates when allowing only for single-qubit and $X_i X_j+Y_i Y_j$ two-qubit interactions, and in the less restrictive setting of allowing for all $X_i X_j$ and $Y_i Y_j$ interactions.
The results are shown in~\cref{fig:fredkin_XY,fig:fredkin_XX,fig:toffoli_XX,fig:toffoli_XY}.
For the Fredkin gate, in the more restrictive $XX$ interactions setting, the biggest fidelity obtained was $\F\simeq 0.94$, while when allowing for all $XX$ and $YY$ interactions the maximum fidelity obtained was $\F\simeq0.999$.
For the Toffoli gate, the maximum fidelity obtained in the $XX$ scenario was $\F\simeq0.94$ as well, while when allowing for all $XX$ and $YY$ interactions the best training results corresponded to $\F\simeq 0.98$.
To have more consistent results, in all the training instances shown here all the hyperparameters, except for the interaction parameters' initial values, were chosen to have the same value.
In particular, each training instance was run for $200$ epochs, each one using $200$ random quantum states as inputs, divided in batches of $2$ elements.
This choice of hyperparameters is mostly empirical, and it is possible for different values to provide better results.

The above provides further evidence for the flexibility of the supervised learning approach, which can produce solutions with good fidelities even in more restrictive scenarios, closer to the capabilities of state of the art experimental architectures.
Furthermore, the values of the interaction strengths for many of the presented solutions are found to be compatible with the capabilities of state of the art circuit-QED architectures with gate times of the order of tens of nanoseconds~\cite{potocnik2018studying}.

Additional solutions and data, as well as the code used to produce them, is available in the GitHub repository
\href{https://github.com/lucainnocenti/quantum-gate-learning-1803.07119}{lucainnocenti/quantum-gate-learning-1803.07119}.
This repository contains all the code used to reproduce the solutions presented in this paper, as well as to train arbitrary gates on arbitrary numbers of qubits.
Even more generally, arbitrary (linearly) parametrised matrices can be used as training model, allowing a high degree of flexibility.

\begin{figure*}[]
	\centering
	\begin{tikzpicture}
		\node[anchor=south west] (A) at (0, 0)%
			{\includegraphics[width=.8\linewidth]{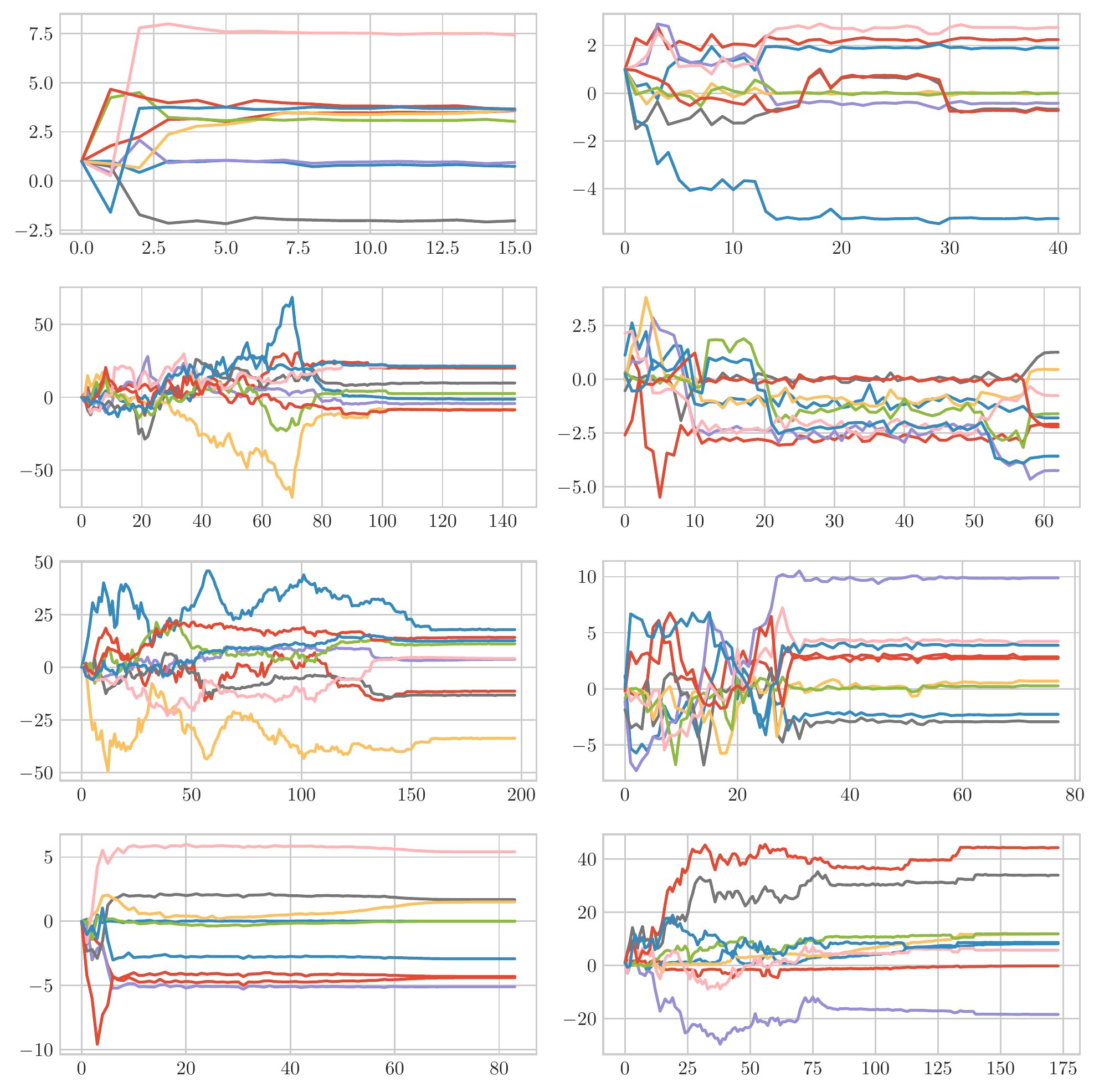}};
		\node[above] at (4., -.2) {$t_e$};
		\node[above] at (11.5, -.2) {$t_e$};
	\end{tikzpicture}%
	\caption{
		Training histories for the \textbf{Toffoli} gate with only diagonal interactions.
		In each plot are reported the values of the $9$ network parameters during the training process, for each training epoch $t_e$.
		Each training process was left running until convergence to unit fidelity, therefore, the number of epochs in the horizontal axes differs for different trainings instances.
		The histories shown here correspond to training instances in which the parameters were initialised at various values, as seen from the leftmost values in each plot.
	}
	\label{fig:toffoli_diagonal_parhistories}
\end{figure*}

\begin{figure*}[]
	\centering
		\begin{tikzpicture}
		\node[anchor=south west] (A) at (0, 0)%
			{\includegraphics[width=.8\linewidth]{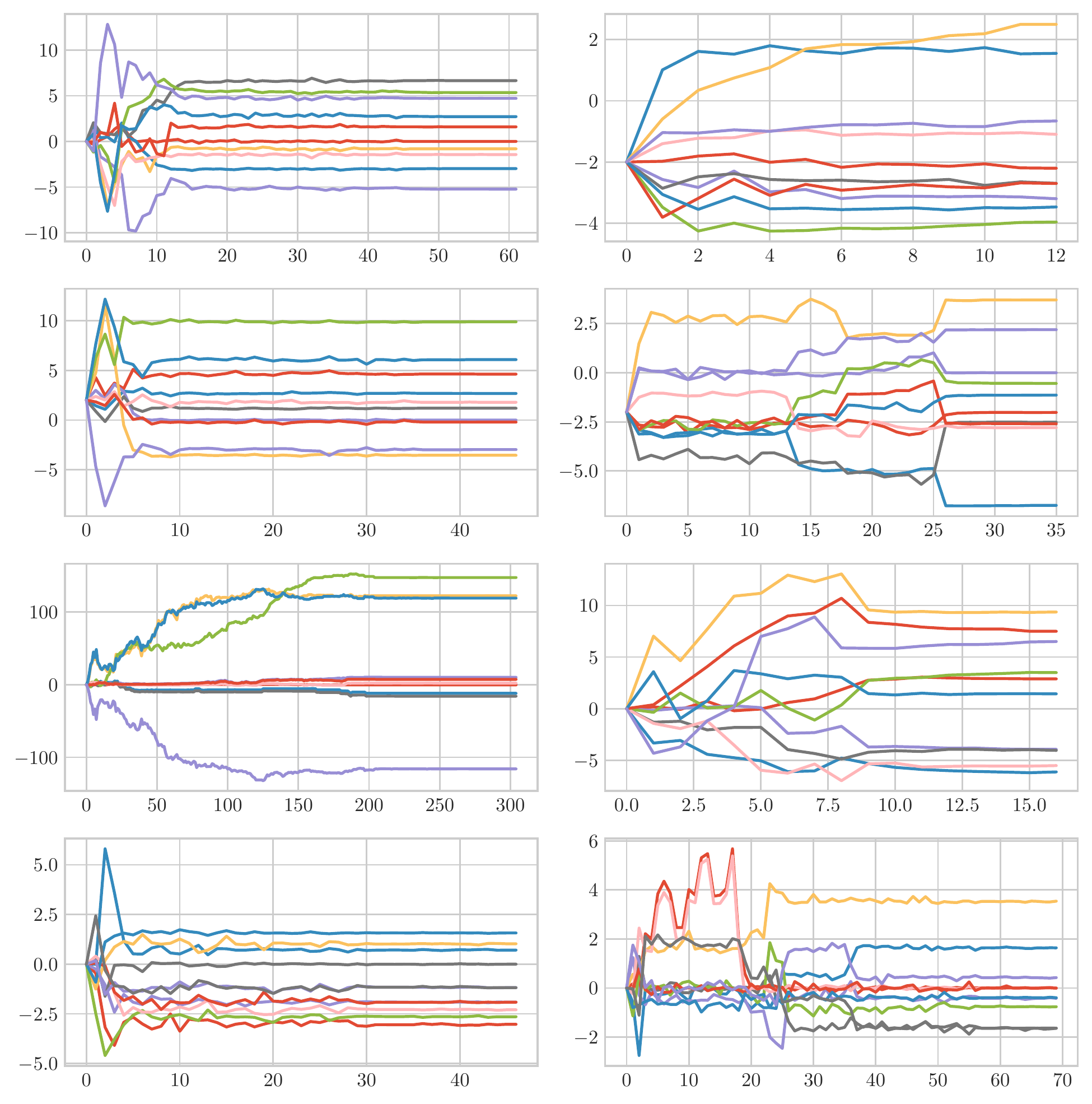}};
		\node[above] at (4., -.2) {$t_e$};
		\node[above] at (11.5, -.2) {$t_e$};
	\end{tikzpicture}%
	\caption{
		Training histories for the \textbf{Fredkin} gate with only diagonal interactions.
		In each plot are reported the values of the $9$ network parameters during the training process, for each training epoch $t_e$.
		Each training process was left running until convergence to unit fidelity, therefore, the number of epochs in the horizontal axes differs for different trainings instances.
		The histories shown here correspond to training instances in which the parameters were initialised at various values, as seen from the leftmost values in each plot.
	}
	\label{fig:fredkin_diagonal_parhistories}
\end{figure*}

\begin{figure*}[]
	\centering
	\begin{tikzpicture}
		\node[anchor=south west] (A) at (0, 0)%
			{\includegraphics[width=.8\linewidth]{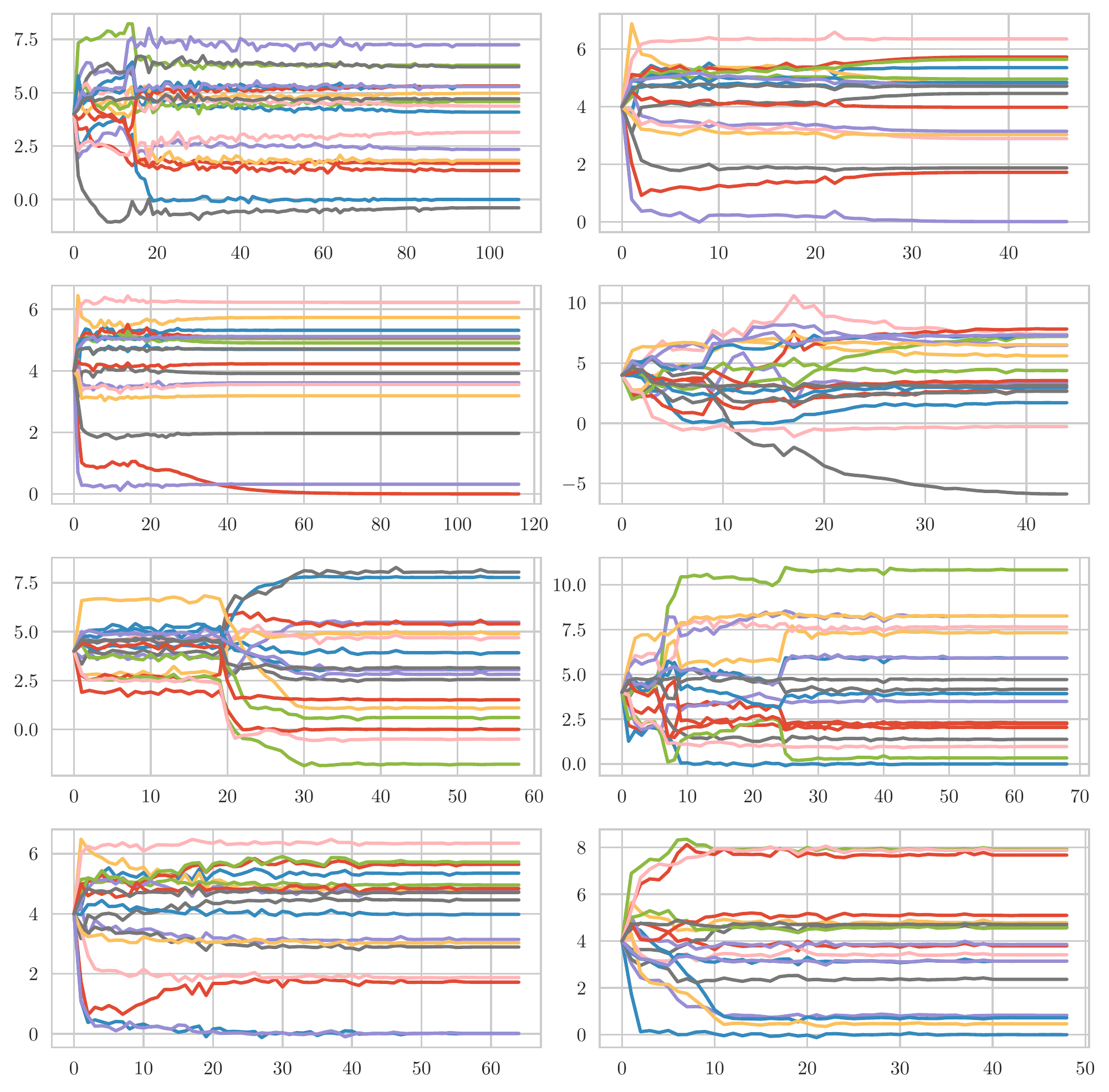}};
		\node[above] at (4., -.2) {$t_e$};
		\node[above] at (11.5, -.2) {$t_e$};
	\end{tikzpicture}%
	\caption{
		Training histories for the \textbf{double-Fredkin} gate with only diagonal interactions.
		In each plot are reported the values of the $18$ network parameters during the training process, for each training epoch $t_e$.
		Each training process was left running until convergence to unit fidelity, therefore, the number of epochs in the horizontal axes differs for different trainings instances.
		All the histories shown here correspond to training instances in which the parameters were initialised to $4$.
	}
	\label{fig:doublefredkin_diagonal_parhistories}
\end{figure*}

\newcommand{\graphicswithlabel}[3]{%
	\begin{tikzpicture}
		\node[anchor=north west] (A) at (0, 0)%
			{\includegraphics[width=.48\linewidth]{#2}};
		\node[above] at (1.4, -.85) {\small\textbf{(#1)}};
		\node[above] at (4.6, -4.65) {#3};
		\node[above] at (1., -.2) {$\F_\bslambda(\psi)$};
	\end{tikzpicture}%
}
\begin{figure*}
	\centering
	\graphicswithlabel{a}{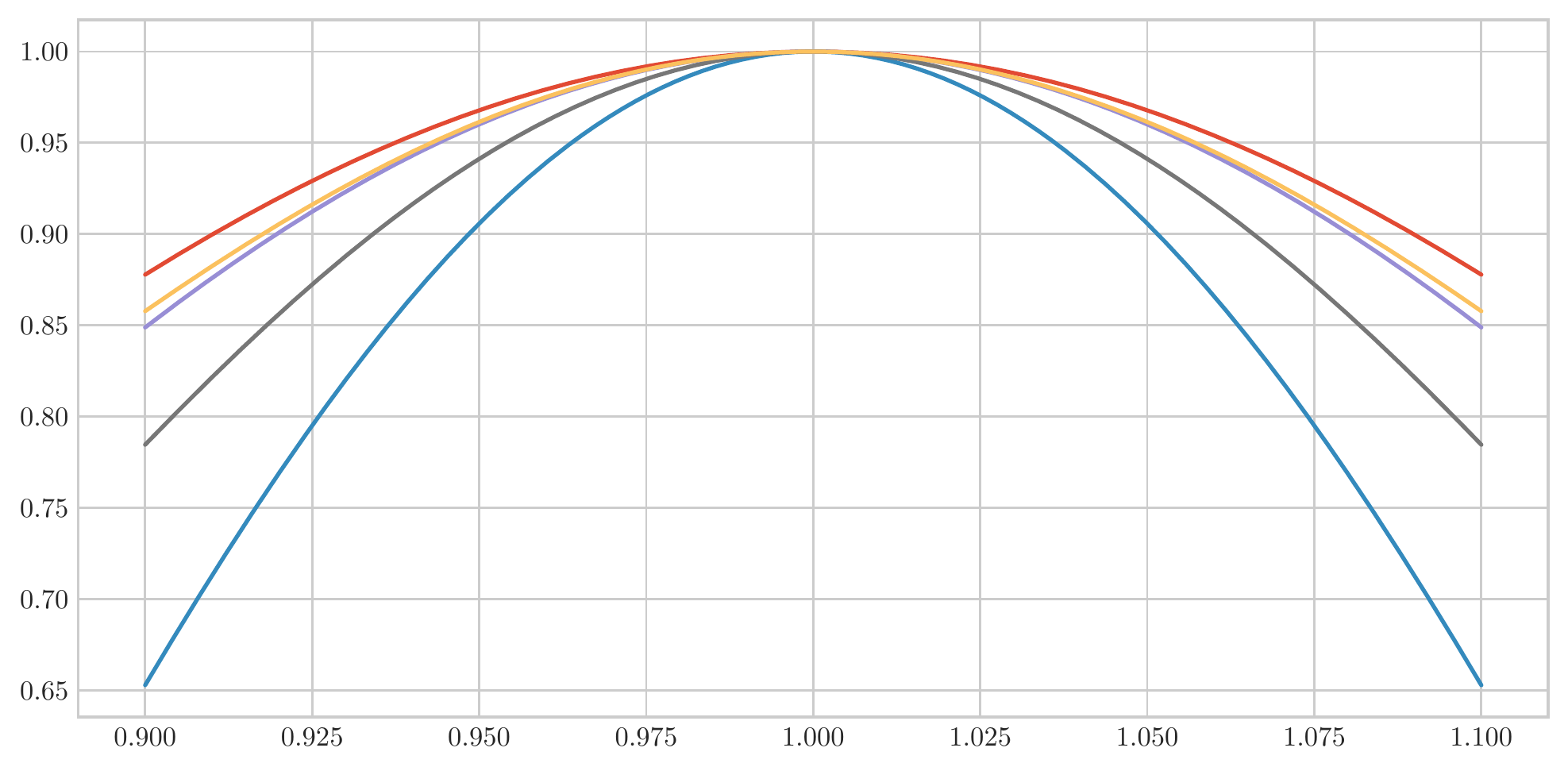}{$\alpha$}\vspace{-6pt}
	\graphicswithlabel{b}{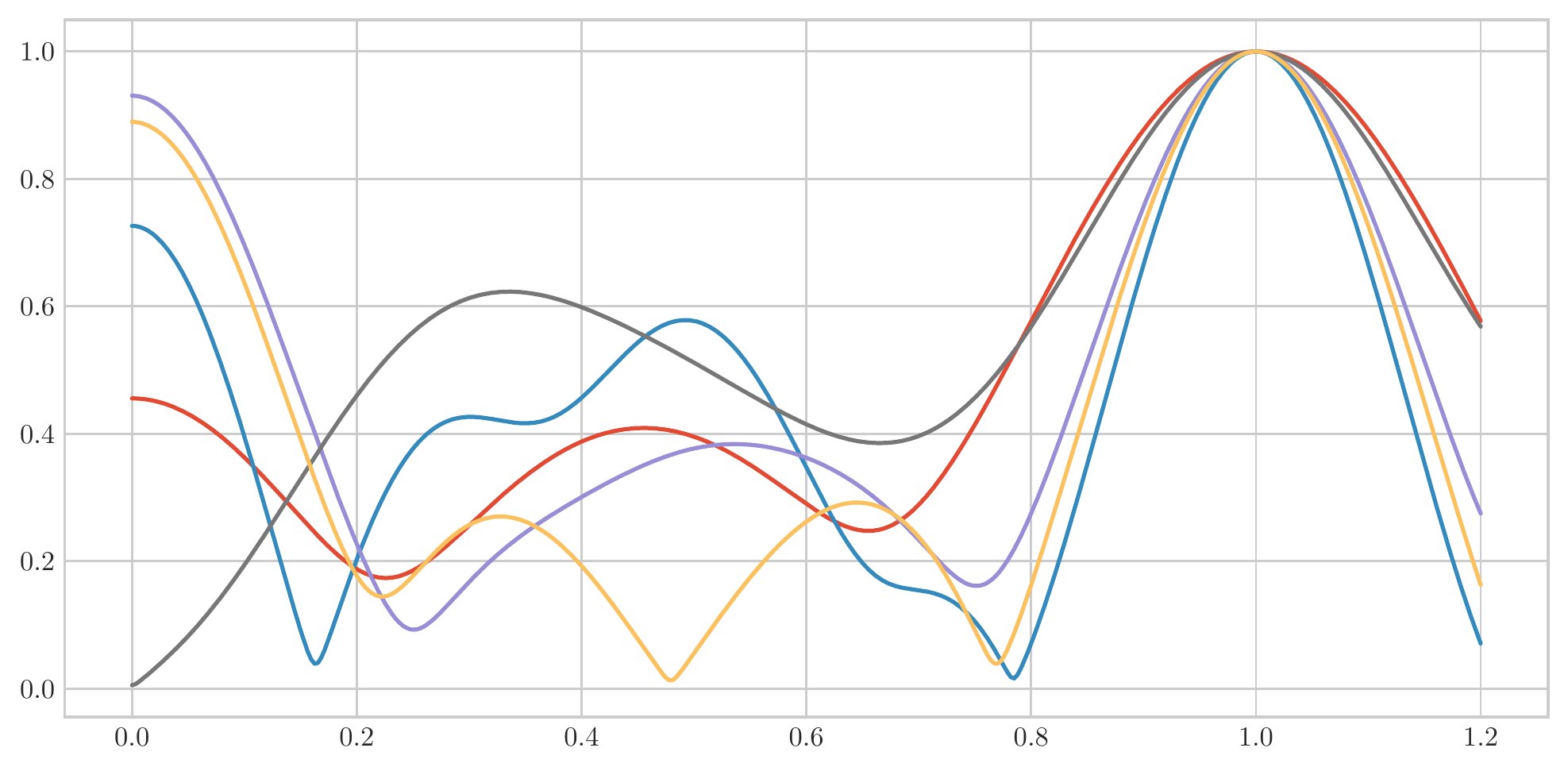}{$\alpha$}
	\graphicswithlabel{c}{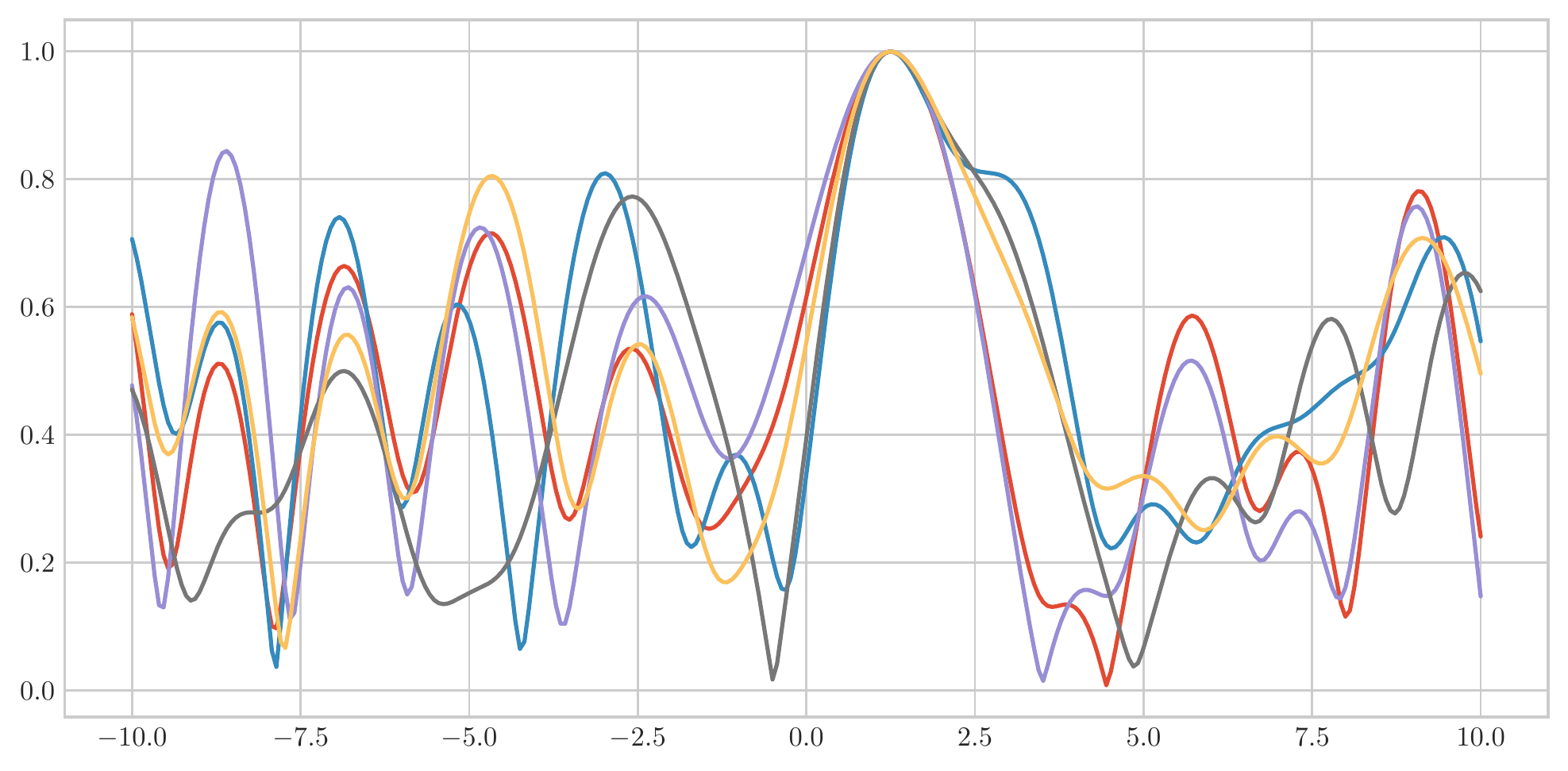}{$\lambda_1$}\vspace{-6pt}
	\graphicswithlabel{d}{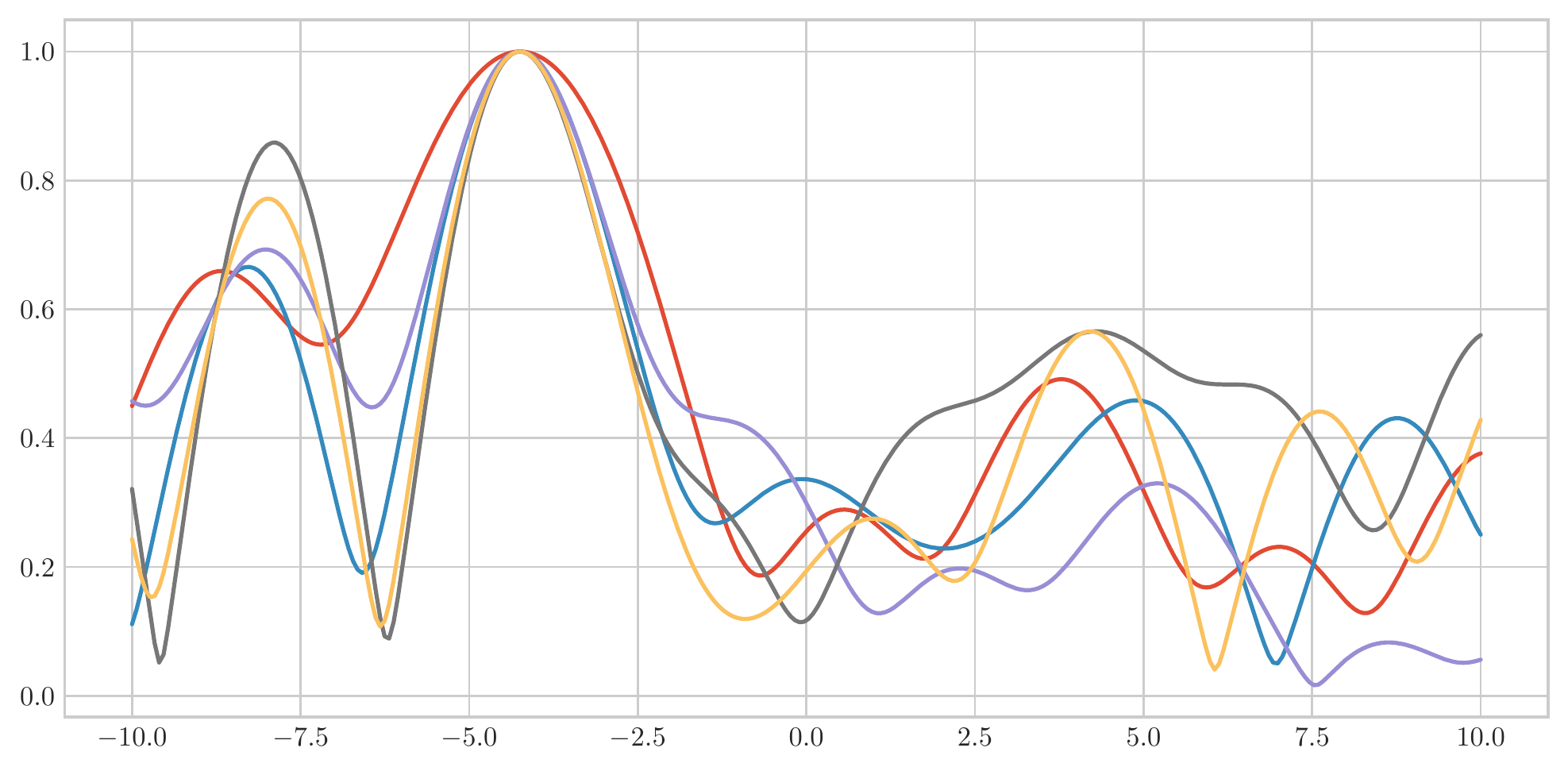}{$\lambda_2$}
	\caption{
		Fidelity $\F_\bslambda(\psi)$ vs variations of $\bslambda$, for different test states, for the \textbf{Toffoli} gate. The five test states $\psi$ are sampled randomly.
		\textbf{(a)} Global relative variations of $\bslambda$, that is, plotting the fidelity against $\alpha\bslambda$ for $0.9 \le\alpha\le 1.1$.
		Note that this is equivalent to studying how the fidelity changes with respect to uncertainties in the evolution time, that is, how much does $\exp(i\HH t')$ differ from $\exp(i\HH t)$.
		\textbf{(b)} Same as \emph{(a)} but with $0\le\alpha\le 1.2$.
		\textbf{(c)} Plot of $\F_\bslambda(\psi)$ against \emph{absolute} variations of a single element of $\bslambda$, in this case the first one, i.e. we take $\lambda_1\in[-10,10]$.
		\textbf{(d)} Like \emph{(c)} but for $\lambda_2$.
	}
	\label{fig:toffoli_fidVSpars}
\end{figure*}

\renewcommand{\graphicswithlabel}[3]{%
	\begin{tikzpicture}
		\node[anchor=north west] (A) at (0, 0)%
			{\includegraphics[width=.48\linewidth]{#2}};
		\node[above] at (1.4, -.85) {\small\textbf{(#1)}};
		\node[above] at (4.6, -4.65) {#3};
		\node[above] at (1., -.2) {$\F_\bslambda(\psi)$};
	\end{tikzpicture}%
}
\begin{figure*}
	\centering
	\graphicswithlabel{a}{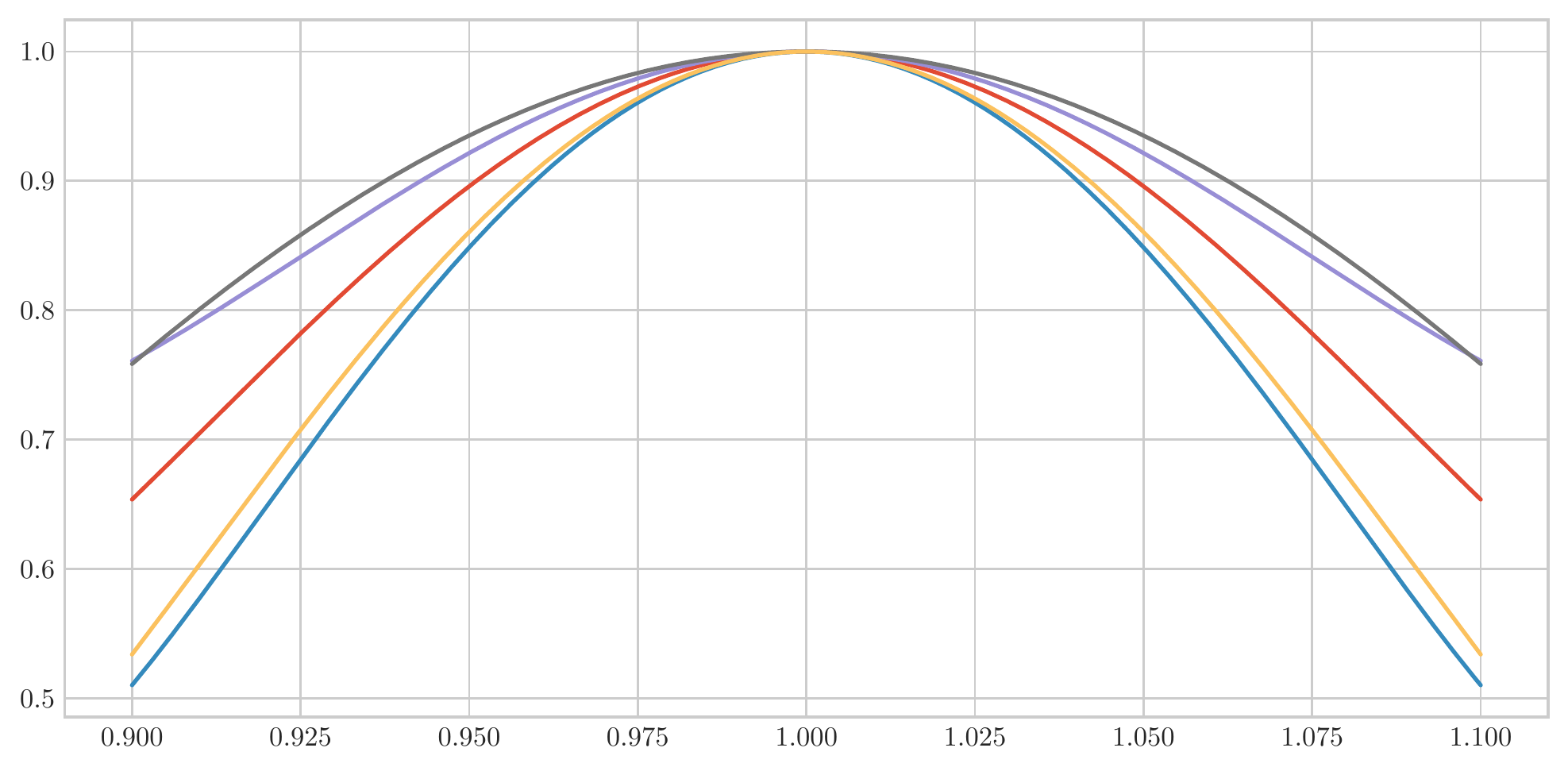}{$\alpha$}\vspace{-6pt}
	\graphicswithlabel{b}{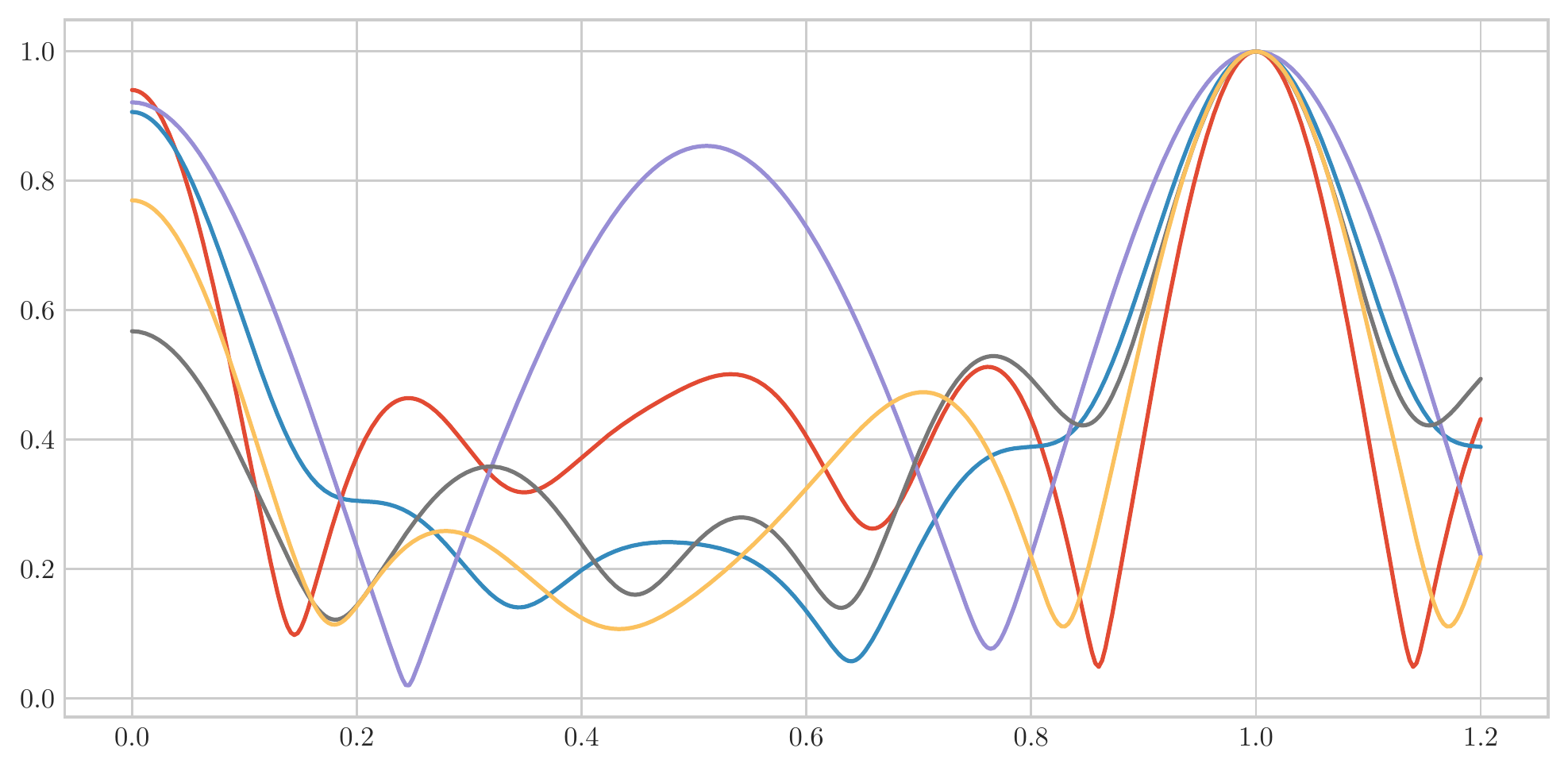}{$\alpha$}
	\graphicswithlabel{c}{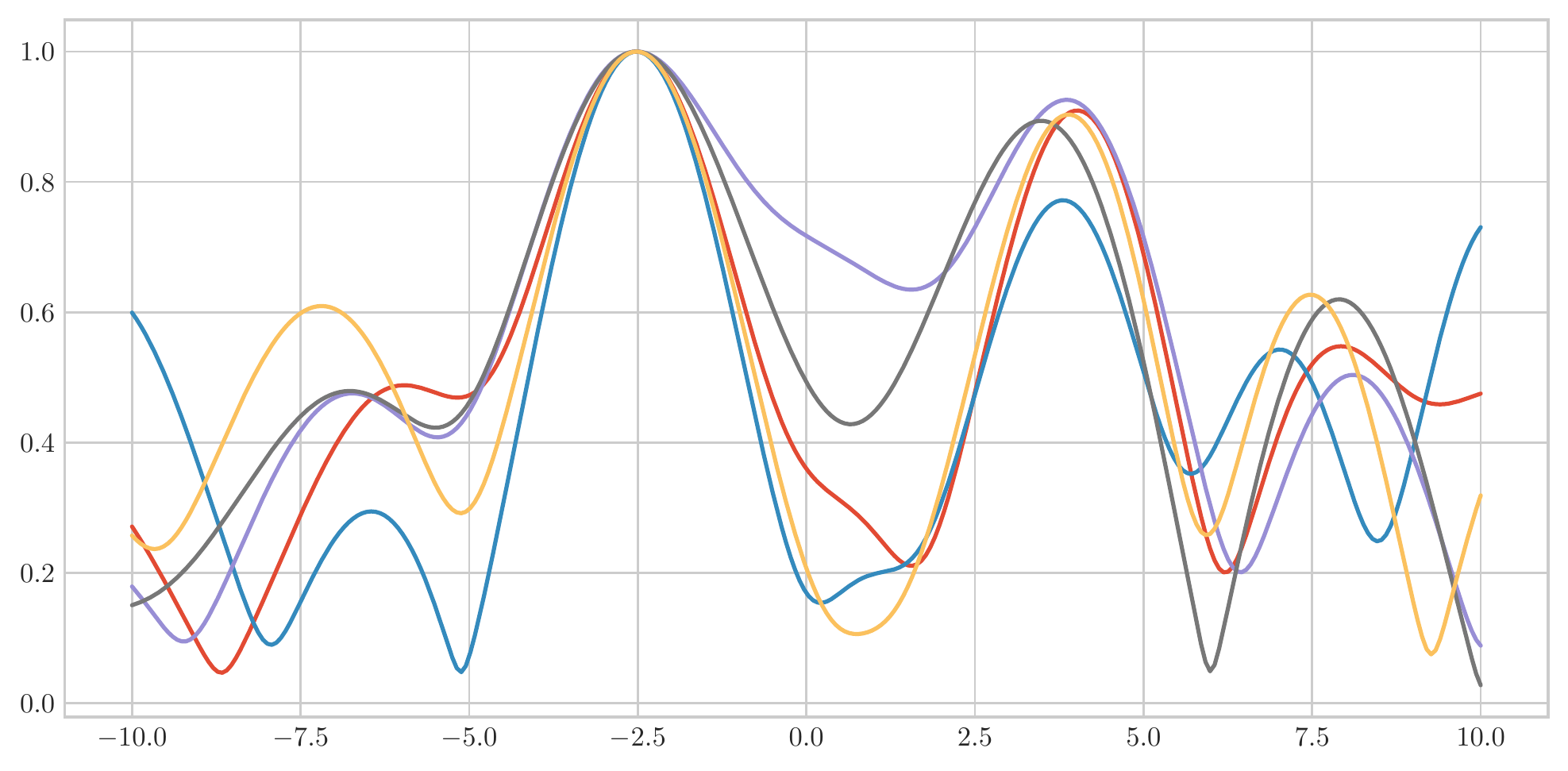}{$\lambda_1$}\vspace{-6pt}
	\graphicswithlabel{d}{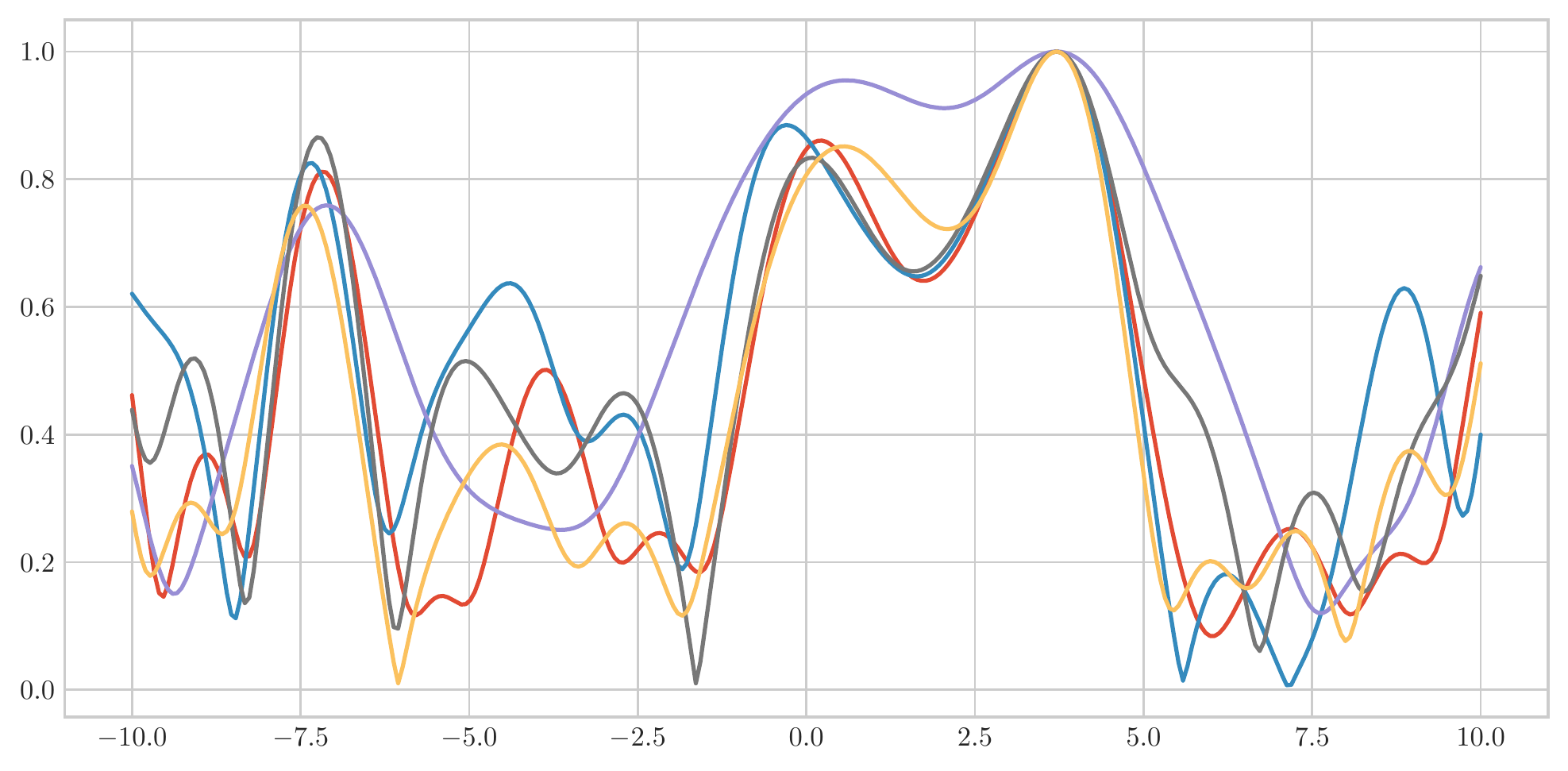}{$\lambda_2$}
	\caption{
		Fidelity $\F_\bslambda(\psi)$ vs variations of $\bslambda$, for different test states, for the \textbf{Fredkin} gate. The five test states $\psi$ are sampled randomly.
		\textbf{(a)} Global relative variations of $\bslambda$, that is, plotting the fidelity against $\alpha\bslambda$ for $0.9 \le\alpha\le 1.1$.
		Note that this is equivalent to studying how the fidelity changes with respect to uncertainties in the evolution time, that is, how much does $\exp(i\HH t')$ differ from $\exp(i\HH t)$.
		\textbf{(b)} Same as \emph{(a)} but with $0\le\alpha\le 1.2$.
		\textbf{(c)} Plot of $\F_\bslambda(\psi)$ against \emph{absolute} variations of a single element of $\bslambda$, in this case the first one, i.e. we take $\lambda_1\in[-10,10]$.
		\textbf{(d)} Like \emph{(c)} but for $\lambda_2$.
	}
	\label{fig:fredkin_fidVSpars}
\end{figure*}

\renewcommand{\graphicswithlabel}[3]{%
	\begin{tikzpicture}
		\node[anchor=north west] (A) at (0, 0)%
			{\includegraphics[width=.48\linewidth]{#2}};
		\node[above] at (1.4, -.85) {\small\textbf{(#1)}};
		\node[above] at (4.6, -4.65) {#3};
		\node[above] at (1., -.2) {$\F_\bslambda(\psi)$};
	\end{tikzpicture}%
}
\begin{figure*}
	\centering
	\graphicswithlabel{a}{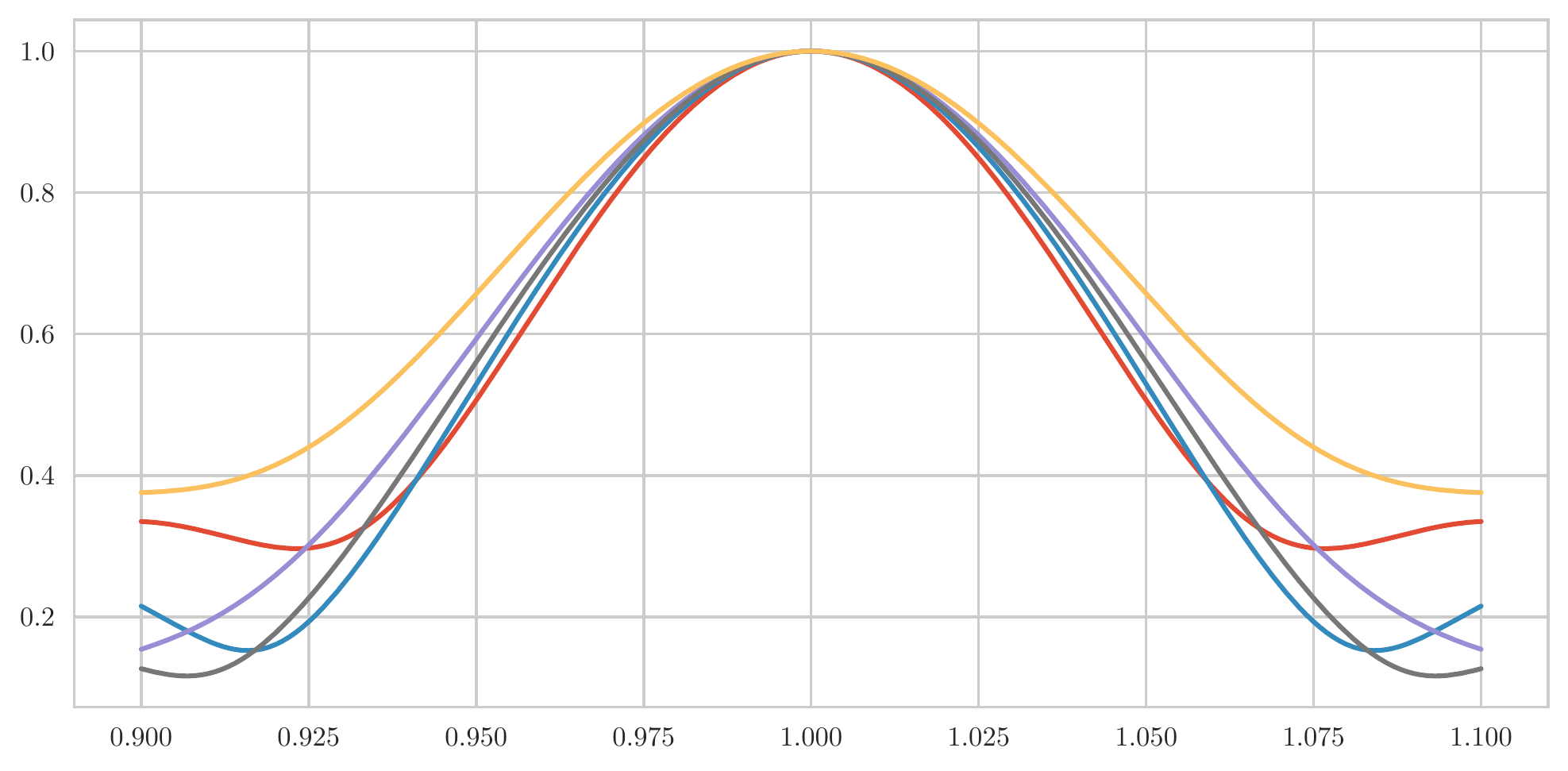}{$\alpha$}\vspace{-6pt}
	\graphicswithlabel{b}{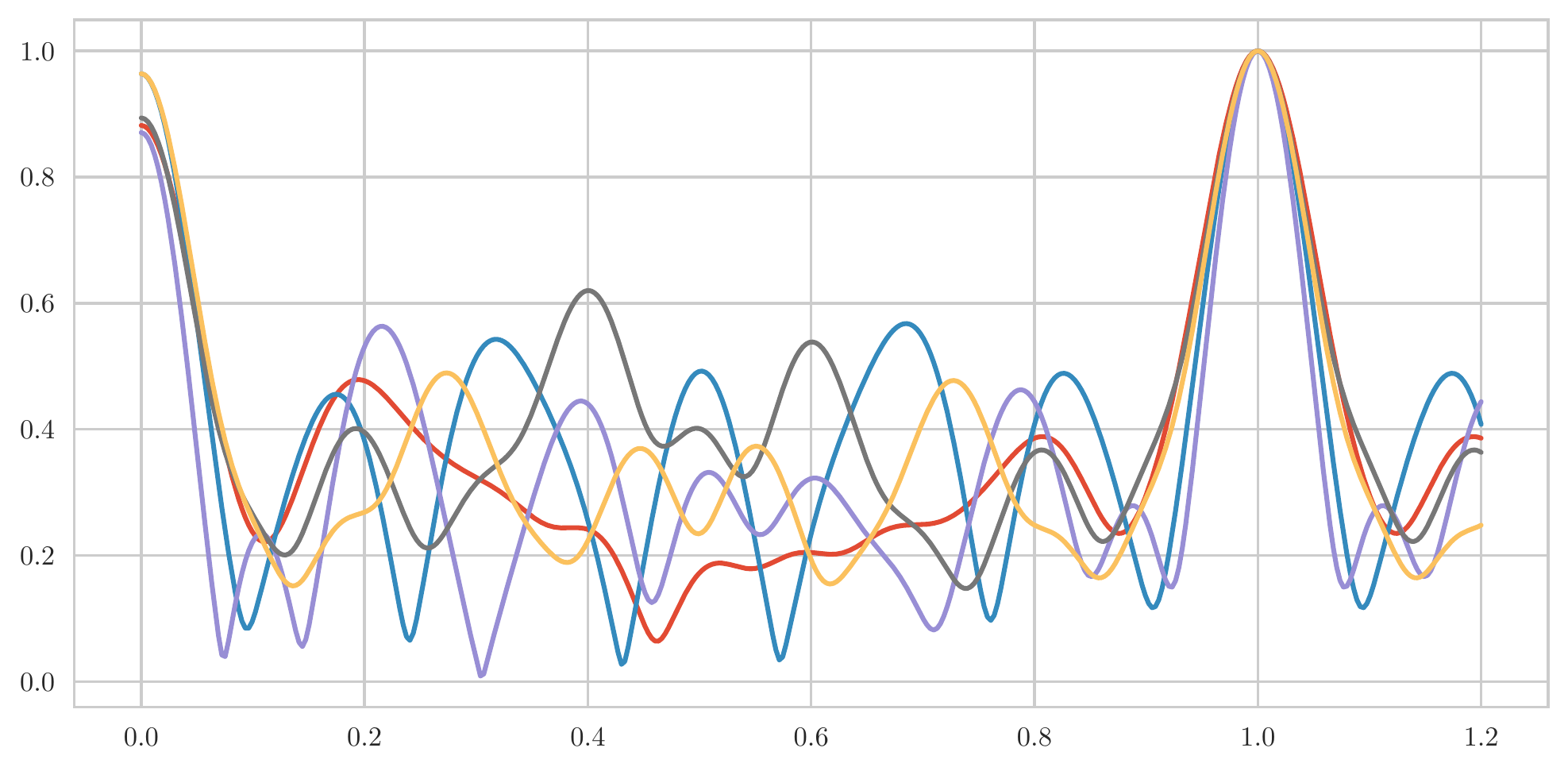}{$\alpha$}
	\graphicswithlabel{c}{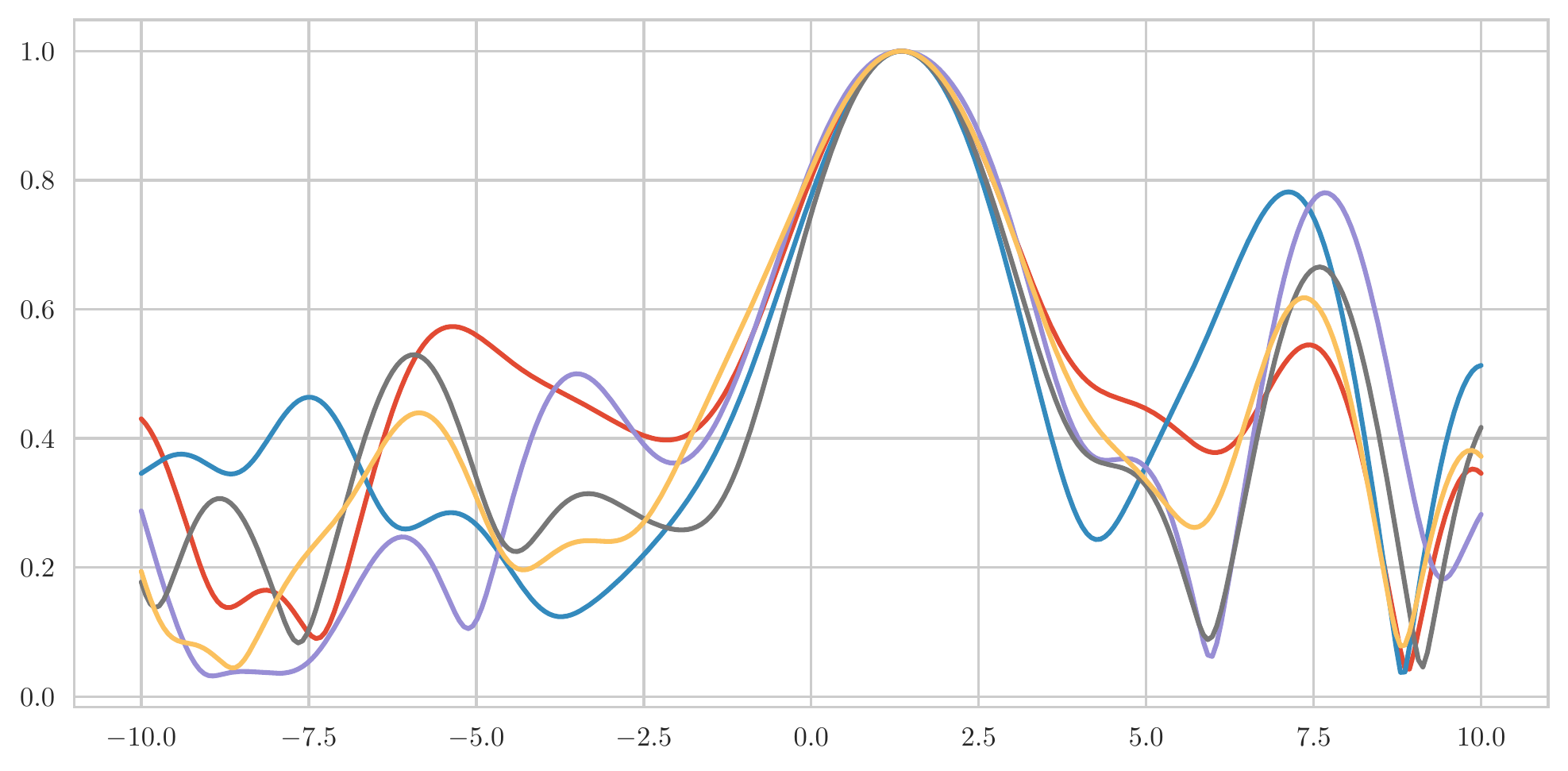}{$\lambda_1$}\vspace{-6pt}
	\graphicswithlabel{d}{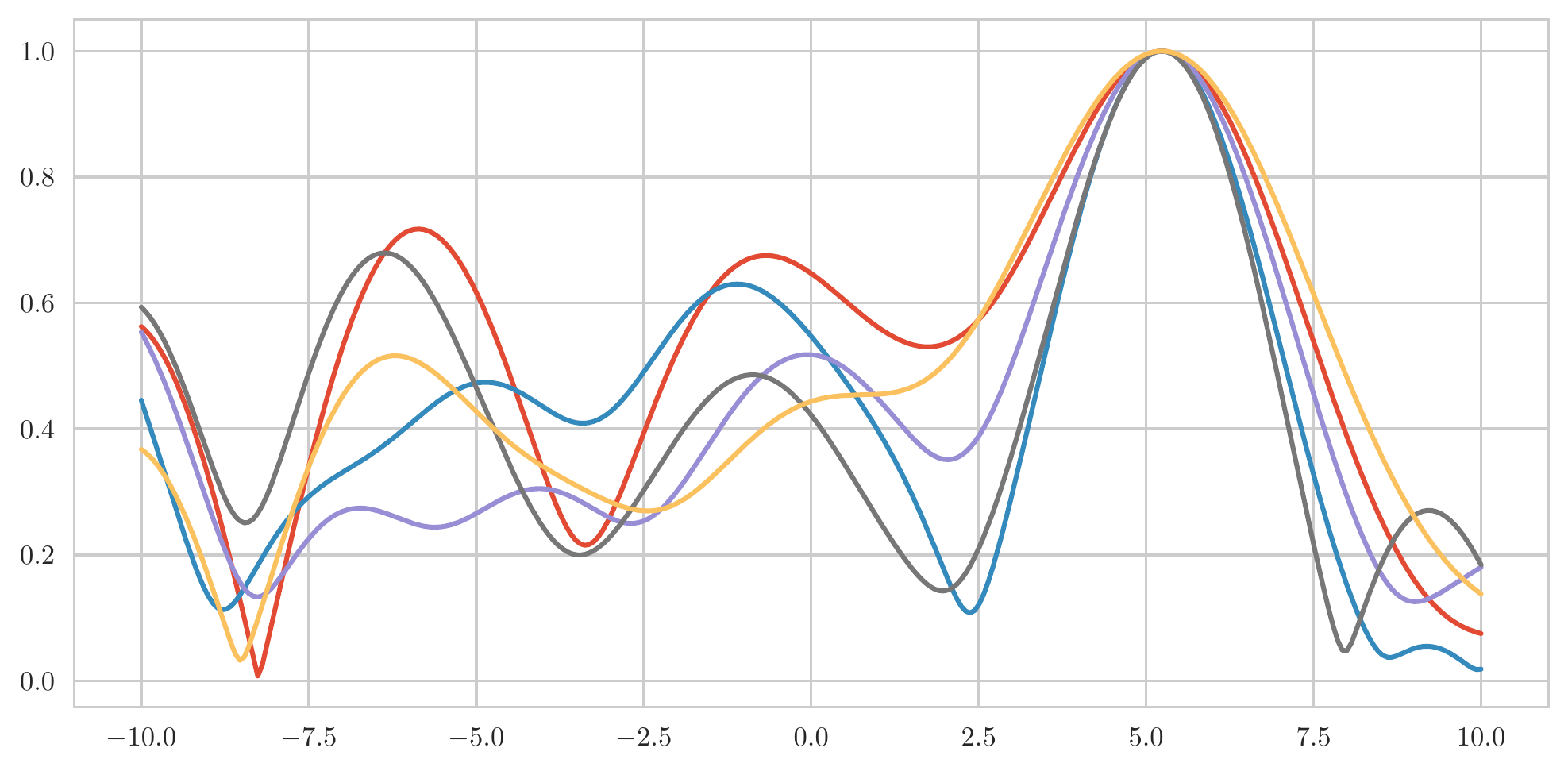}{$\lambda_2$}
	\caption{
		Fidelity $\F_\bslambda(\psi)$ vs variations of $\bslambda$, for different test states, for the \textbf{double Fredkin} gate. The five test states $\psi$ are sampled randomly.
		\textbf{(a)} Global relative variations of $\bslambda$, that is, plotting the fidelity against $\alpha\bslambda$ for $0.9 \le\alpha\le 1.1$.
		Note that this is equivalent to studying how the fidelity changes with respect to uncertainties in the evolution time, that is, how much does $\exp(i\HH t')$ differ from $\exp(i\HH t)$.
		\textbf{(b)} Same as \emph{(a)} but with $0\le\alpha\le 1.2$.
		\textbf{(c)} Plot of $\F_\bslambda(\psi)$ against \emph{absolute} variations of a single element of $\bslambda$, in this case the first one, i.e. we take $\lambda_1\in[-10,10]$.
		\textbf{(d)} Like \emph{(c)} but for $\lambda_2$.
	}
	\label{fig:doublefredkin_fidVSpars}
\end{figure*}

\newcommand{\traininggraphicswithlabel}[1]{%
	\begin{tikzpicture}
		\node[anchor=north west] (A) at (0, 0)%
			{\includegraphics[width=.85\linewidth]{#1}};
		\node[above] at (0, -.8) {$\barF$};
	\end{tikzpicture}%
}

\begin{figure*}
	\centering
	\traininggraphicswithlabel{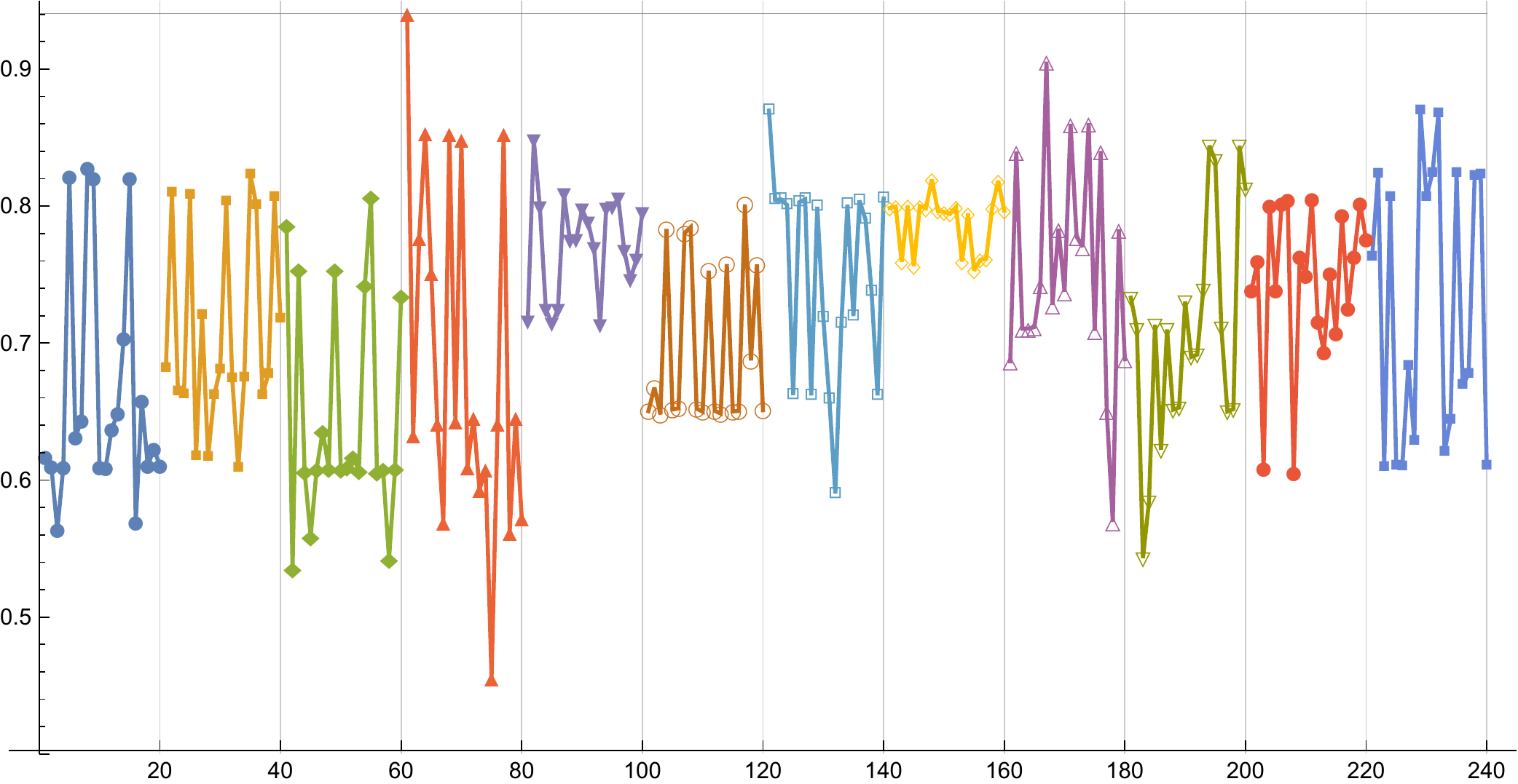}
	\caption{
		Training results for the Fredkin gate with a generator containing one-qubit interactions and two-qubit interactions of the form $J_{ij}(X_i X_j + Y_i Y_j)$.
		Every point shows the final fidelity obtained at the end of a training procedure.
		The hyperparameters, as well as the total number of training iterations, are kept the same in all the training instances shown here.
		The initial parameters' values are the same within each horizontal sector, but changed between different sectors.
		The initial parameters' values within each sector have been chosen as all equal to $c$ (that is, $\lambda_i=c$ for all $i$). The values of $c$ are $0, 1,..., 10$, with $c=0$ in the leftmost sector and $c=10$ in the last to rightmost one.
		The rightmost sector contains the results of training attempts with the initial values chosen at random (that is, with $\lambda_i$ sampled according to the uniform normal distribution, independently for each $i$).
		The greatest reached fidelity, obtained with initial parameters' values $\lambda_i=3$, is $\F\simeq0.94$.
	}
	\label{fig:fredkin_XX}
\end{figure*}

\begin{figure*}
	\centering
	\traininggraphicswithlabel{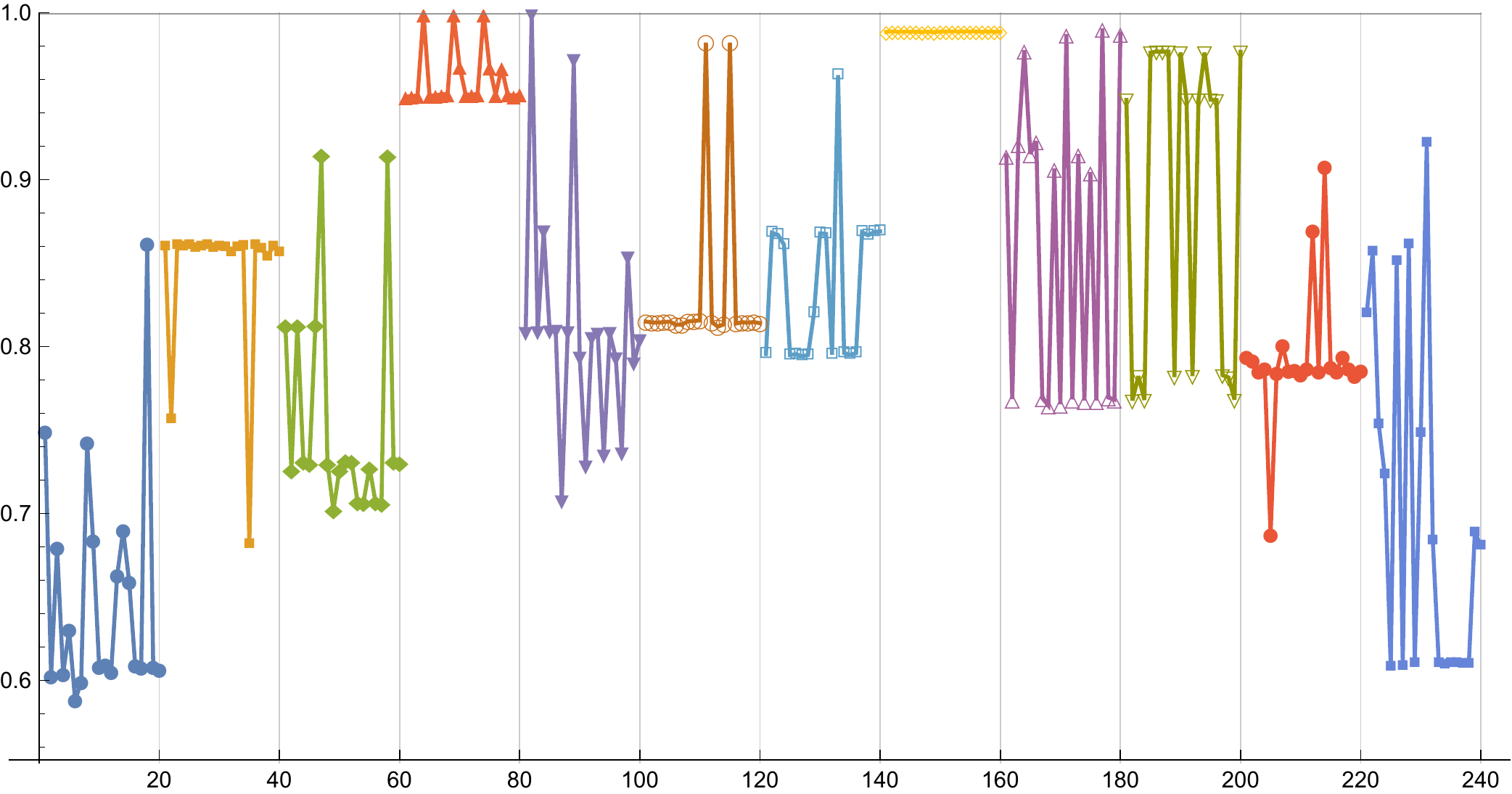}
	\caption{
		Training results for the Fredkin gate with a generator containing all one-qubit interactions and two-qubit interactions of the form $J^{(1)}_{ij}X_i X_j + J^{(2)}_{ij}Y_i Y_j$.
		The initial conditions are chosen as in~\cref{fig:fredkin_XX}.
		The maximum fidelities obtained are $\F\simeq0.999$, obtained in multiple instances $c=3$ and $c=4$ sectors.
	}
	\label{fig:fredkin_XY}
\end{figure*}

\begin{figure*}
	\centering
	\traininggraphicswithlabel{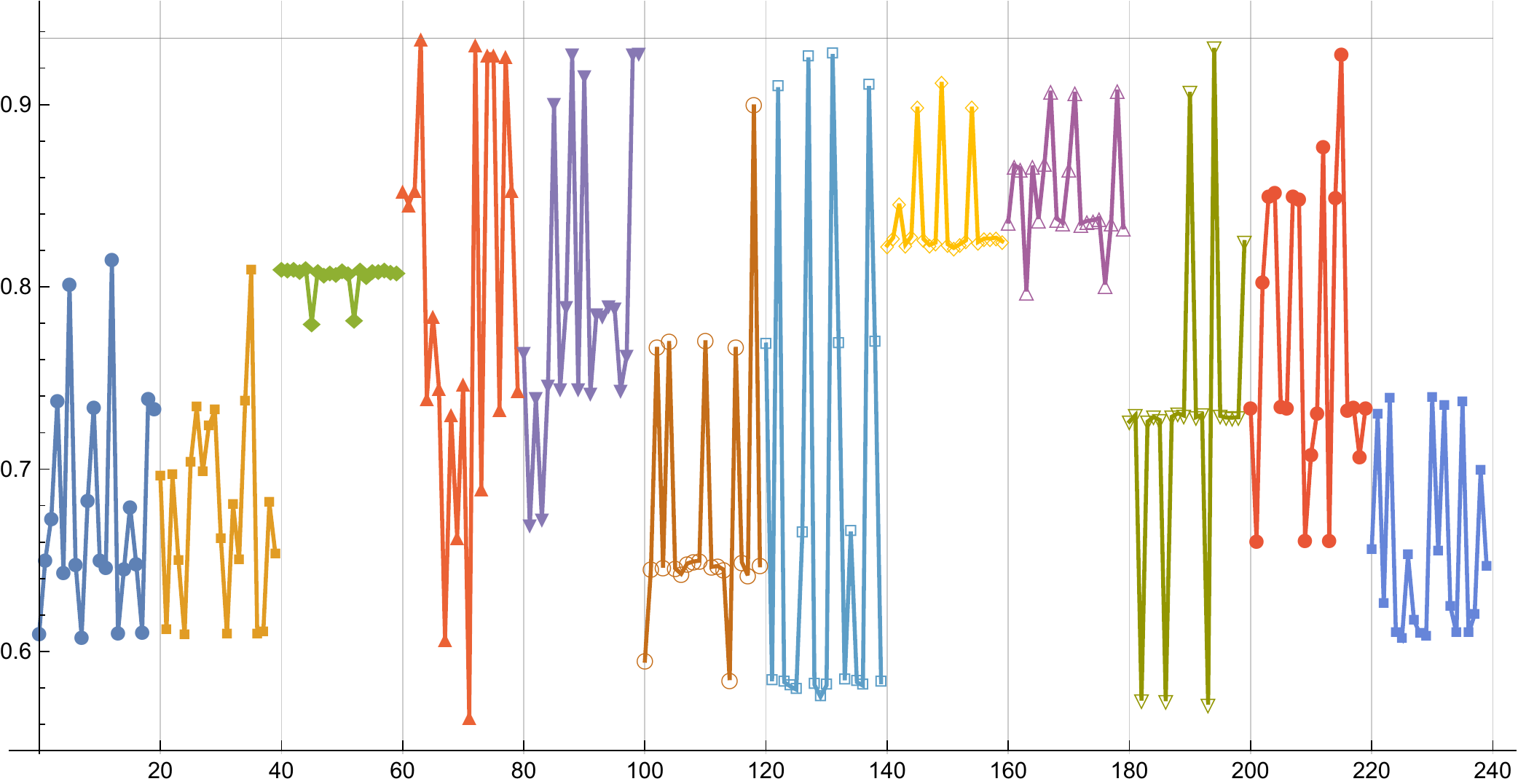}
	\caption{
		Training results for the Toffoli gate with a generator containing all one-qubit interactions and two-qubit interactions of the form $J_{ij}(X_i X_j + Y_i Y_j)$.
		The initial conditions are chosen as in~\cref{fig:fredkin_XX}.
		The maximum fidelities obtained are $\F\simeq0.94$, obtained in the $c=3$ sector.
	}
	\label{fig:toffoli_XX}
\end{figure*}

\begin{figure*}
	\centering
	\traininggraphicswithlabel{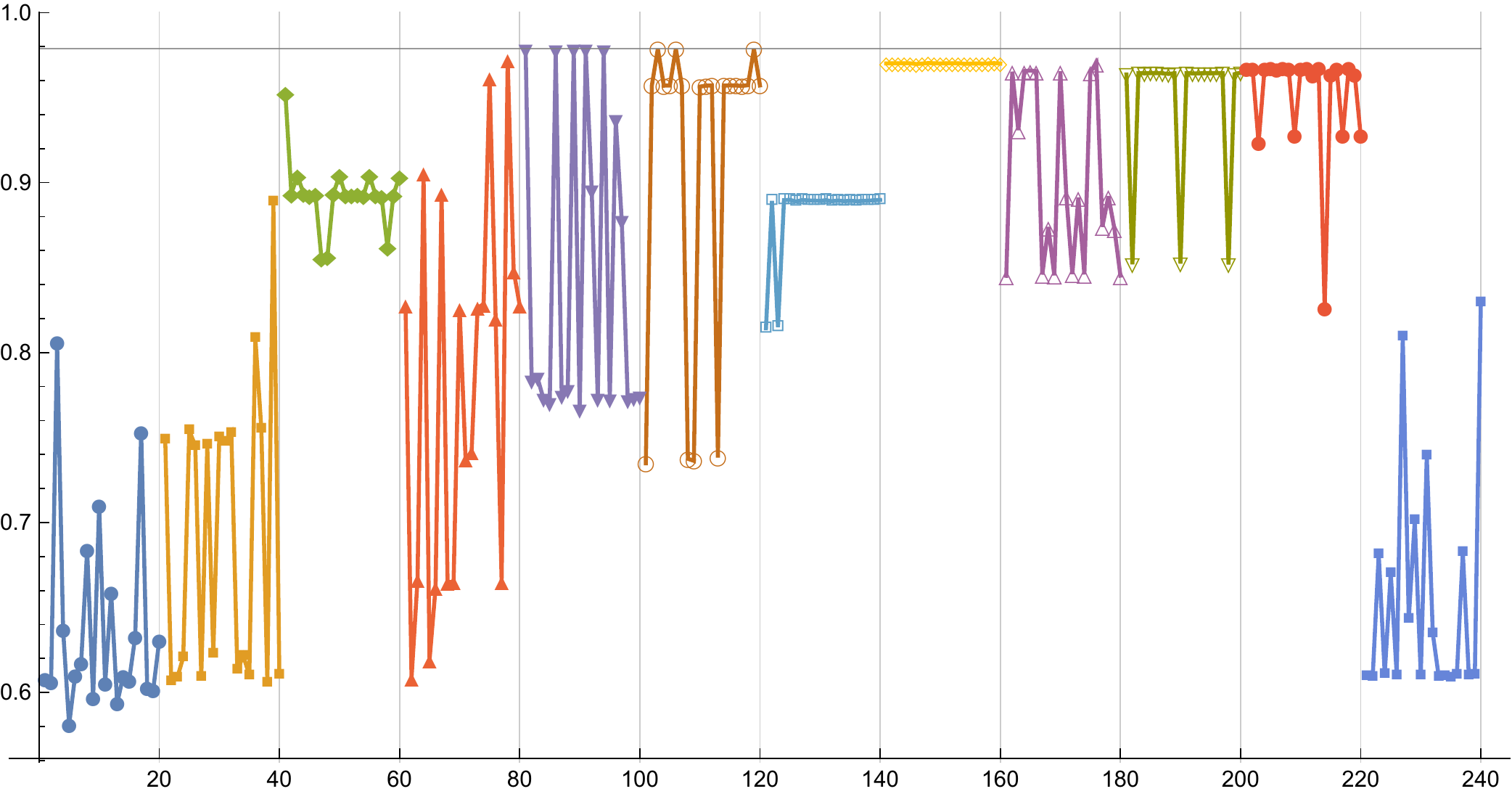}
	\caption{
		Training results for the Toffoli gate with a generator containing all one-qubit interactions and two-qubit interactions of the form $J^{(1)}_{ij}X_i X_j + J^{(2)}_{ij}Y_i Y_j$.
		The initial conditions are chosen as in~\cref{fig:fredkin_XX}.
		The maximum fidelities obtained are $\F\simeq0.98$, obtained in the $c=4$ and $c=5$ sectors.
	}
	\label{fig:toffoli_XY}
\end{figure*}

\end{document}